# Pressure-driven viscoelastic flow in axisymmetric geometries with application to the hyperbolic pipe

**Kostas D. Housiadas**

*Department of Mathematics, University of the Aegean, Karlovassi, Samos, 83200, Greece*

**Antony N. Beris**

*Department of Chemical and Biomolecular Engineering, University of Delaware, Newark, Delaware 19716, USA*

**Abstract**

We investigate theoretically the steady incompressible viscoelastic flow in a rigid axisymmetric tube (cylindrical pipe) with varying cross-section. We use the Oldroyd-B viscoelastic constitutive equation to model the fluid viscoelasticity. First, we derive new exact results expressed in the form of general formulas: for the average pressure-drop through the pipe as a function of the wall shear rate and the viscoelastic axial normal extra-stress, for the viscoelastic extra-stress tensor at the axis of symmetry and the Trouton ratio of the fluid as function of the fluid velocity at the axis, and for the viscoelastic extra-stress tensor along the wall in terms of the tangential shear rate at the wall. We then proceed by exploiting the classic lubrication approximation, valid for small values of the square of the aspect ratio of the pipe, to simplify the original governing equations. The final equations are solved analytically using a regular perturbation scheme in terms of the Deborah number, *De*, up to eight order in *De*. For the specific case of a hyperbolic pipe, we reveal that the pressure-drop and the Trouton ratio (when reduced by their corresponding Newtonian values) can be recast in terms of a modified Deborah number, $De_m$, and the polymer viscosity ratio, $\eta$, only. Furthermore, we enhance the convergence and accuracy of the eight-order solutions by deriving transformed analytical formulas using Padé diagonal approximants. The results show the decrease of the pressure drop and the enhancement of the Trouton ratio with increasing $De_m$ and/or increasing $\eta$. Comparison of the transformed solutions with numerical simulations of the lubrication equations using pseudospectral methods shows excellent agreement between the results revealing the robustness, validity and efficiency of the theoretical methods and techniques developed in this work.

1. **Introduction**

Pressure-driven flows of viscoelastic fluids in narrow and long tubes (channel or pipes) are widely encountered in industrial processes, such as extrusion (Pearson 1985; Tadmor & Gogos 2013), in applications such as microfluidic extensional rheometers (Ober *et al.* 2013), in devices for subcutaneous drug administration (Allmendinger *et al.* 2014; Fischer *et al.* 2015) and many others. The complex rheological behavior of viscoelastic fluids affects various features and properties of these flows, such as the average pressure drop along the tube, $\Delta\Pi^*$, as function of the flow rate, $Q^*$, and the elongational viscosity of the fluid, $\eta_{el}^*$, as function of the extensional rate; throughout the paper a star superscript denotes dimensional quantity.

Contraction flows, and especially flows through hyperbolic geometries, have attracted a lot of attention because of the almost constant extensional rate developed along the midplane for planar geometries or along the axis of symmetry for axisymmetric ones. These flows have been used to measure $\eta_{el}^*$ by relating $\Delta\Pi^*$ to $Q^*$. The $\Delta\Pi^* - Q^*$ relationship allows for the extraction of $\eta_{el}^*$ (see e.g., Wang & James 2011; Ober *et al.* 2013; Nyström *et al.* 2016; Kim *et al.* 2018) although it should be emphasized that the measurement of the extensional viscosity (Binding & Jones 1989; James & Walters 1993; Binding & Walters 1998; Oliveira *et al.* 2007; Ober *et al.* 2013; Keshavarz & McKinley 2016) or its theoretical prediction, as for instance presented by the preliminary works of Cogswell (1972, 1978) and further improved by Binding (1988) and later by James (2016), are both challenging because a steady and spatially uniform (i.e., homogeneous) flow from a Lagrangian point of view cannot be achieved easily (Petrie 2006). Other published works on the subject also include the experimental work of Lee & Muller (2017), the simulations of Nyström *et al.* (2012, 2016, 2017) and Feigl *et al.* (2003), and the optimization work with respect to the geometry of Zografos *et al.* 2020, with goal to produce a constant strain-rate along the midplane of a symmetric channel. Note however that some authors have claimed that given that the flow in this geometry is not purely extensional, it is difficult to estimate the resistance to extensional motion directly from the $\Delta\Pi^* - Q^*$ experimental data (Nyström *et al.* 2016; James 2016; Hsiao *et al.* 2017).

For a very long time now, the theoretical study of viscoelastic flows has been proven to be a very demanding and complicated task. This is attributed mainly to the fact that the relevant governing equations which predict the spatial and/or temporal evolution of the viscoelastic extra-stresses due to flow deformation are highly nonlinear. This holds true even when the most basic differential constitutive equation, i.e. the Upper Convected Maxwell (UCM) model, is used, and under the simplest possible flow conditions (steady, laminar and creeping conditions). Thus, the development and use of theories and techniques that further simplify the original governing equations is imperative. Such a theory is the lubrication theory; a simple and efficient asymptotic technique used widely for the modeling of thin fluid films (Langlois 1964; Ockendon & Ockendon 1995; Leal 2007), the motion of particles near surfaces (Goldman *et al.* 1967; Stone 2005), the flow in microchannels with known geometry (Stone *et al.* 2004; Plouraboué *et al.* 2004; Amyot *et al.* 2007; Tavakol *et al.* 2017; Housiadas & Tsangaris 2022, 2023), and generally for the theoretical study of slow flows in narrow and confined geometries.

Even more specifically, application of the lubrication approximation to study viscoelastic lubricants flows in confined with solid walls tubes (mainly channels) has been initiated and performed in the area of tribology by Tichy (1996) who was the first to derive the lubrication equations for the UCM model. Tichy's work was subsequently followed by others such as Zhang, Matar & Craster 2002 who investigated the dynamic spreading of a surfactant on a thin viscoelastic film using the same constitutive model; Li (2014) who investigated the effect of viscoelasticity on the lubrication performance in thin film flows using the Phan-Thien & Tanner (PTT) model, Gamaniel, Dini & Biancofiore (2021) who studied thin-film sliding lubricant flow in the presence of cavitation using the Oldroyd-B model, and Ahmed & Biancofiore (2021, 2023) who studied lubricants viscoelastic flows using the UCM/Oldroyd-B and FENE type models. Also, theoretical investigation of the effect of viscoelasticity, augmented by numerical simulations, on contracting channels with hyperbolic geometry was carried out by Perez-Salas *et al.* (2019) using the PTT model, Boyko & Stone (2022) using the UCM/Oldroyd-B models, and Housiadas & Beris (2023) who advanced the work of Boyko & Stone (2022) by deriving high-order asymptotic solutions (up to eight-order in the perturbation parameter) using the UCM/Oldroyd-B, PTT, Giesekus and FENE-P models. An extended list of literature works for viscoelastic flows in internal and confined geometries (both planar and axisymmetric) has been compiled and can be found in the paper of Boyko & Stone (2022).

Although plenty of computational and experimental works (Oliveira *et al.* 2007; James & Roos 2021; Rothstein & McKinley 1999, 2001; Nigen & Walters 2002; Sousa *et al.* 2009) are available in the literature for axisymmetric geometries, a formal theoretical analysis for the axisymmetric case is still missing. Also, even though the axisymmetric geometry is similar to the planar symmetric case, differences have been observed in experiments (Nigen & Walters 2002; Rodd *et al.* 2005; James & Roos 2021), while the singular mathematical nature of the governing equations at the axis of symmetry of the tube requires special attention and can reveal interesting features of the solution. We also mention that except from the data reported by James & Roos (2021), the experimental data by Rothstein & McKinley (1999, 2001), Nigen & Walters (2002) and Sousa *et al.* (2009) (for 3-D channel flows with square cross-sections though) show an increase in the pressure drop with increasing the fluid viscoelasticity, which is in contrast with the numerical results obtained using macroscopic constitutive models, for instance those of Chilcott & Rallison (1988) using a modified version of the finite-extensibility nonlinear elastic model (FENE-CR), the simulations of Keiller in planar and axisymmetric contractions (1993) using the Oldroyd-B and FENE-P models, the work of Alves *et al.* (2003) using the Oldroyd-B model in planar contraction flow, the work of Aguayo *et al.* (2008) using the Oldroyd-B model, and the work of Aboubacar *et al.* (2002) using the Phan-Thien & Tanner (PTT) model. Those experimental data are in contrast with the theoretical predictions of Boyko & Stone (2022) and Housiadas & Beris (2023) too. However, mesoscopic numerical simulations with the bead–rod and bead–spring models performed by Koppol *et al.* (2009) can somehow resolve this discrepancy. The numerical mesoscopic simulations though are still very demanding and complicated making them prohibited for many researchers; thus, the use of macroscopic constitutive models is the most reasonable approach for the study of viscoelastic flows in complex geometries (for further comments on the discrepancy between the experimental data and the numerical results see Boyko & Stone (2022)).

Here, we undertake a detailed theoretical analysis for the viscoelastic flow in an axisymmetric pipe with variable (non-uniform) cross-section. The wall of the pipe is described with an appropriate smooth, continuous, and adequately differentiable function which we call the "shape function". Note that although the analysis is performed generally in terms of the shape function, we focus on the hyperbolic contracting pipe given its importance to applications. A major goal of our work is to develop a general theoretical framework for the evaluation of the dimensionless elongational viscosity (or Trouton ratio) of the fluid which is not linked to the relationship between $\Delta \Pi^*$ and $Q^*$. This is achieved first by showing that the velocity field at the axis of symmetry of the pipe corresponds to pure uniaxial extension, and then by solving exactly the constitutive model at the axis of symmetry. The analytical solution allows for the development of a general formula for the Trouton ratio in terms of the velocity field only.

Furthermore, we aim to derive approximate analytical solutions for the flow field for this type of flow. To this end, we exploit the lubrication theory to derive a simplified set of governing equations, valid at the limit of a vanishing small aspect ratio of the pipe; the latter is defined as the ratio of the radius of the cross-section at the inlet to the length of the pipe. Furthermore, we investigate the validity and accuracy of the approximate solutions, and we study the effect of the rheological parameters entering in the constitutive model. All these are of fundamental importance for comparison with experimental data and for building reliable mathematical model(s) with good predictive capabilities. We also mention that we are restricted to the UCM/Oldroyd-B models, because as our previous work for the planar hyperbolic geometry showed (Housiadas & Beris 2023), the theoretical predictions under the lubrication approximation using more realistic constitutive models (e.g., the Giesekus, Phan-Thien & Tanner and FENE-P models) are very similar to those for the Oldroyd-B model.

The range of validity and accuracy of the high-order asymptotic expansion in terms of the Deborah number is investigated in three ways (see § 2 for the definition of the Deborah number). First, by employing a method that accelerates the convergence of series such as the diagonal Padé approximants (Padé 1892); this method has been developed and extensively studied in the literature (see, for instance, Baker & Graves-Morris 1981, 1996). The performance of Padé approximants, as well as a few other methods that accelerate the convergence of series, for pure shear and pure elongational flows has been investigated systematically and presented by Housiadas (2017, 2023). These methods have also been applied successfully in a variety of more complex viscoelastic problems (Housiadas 2021, 2023; Housiadas & Beris 2023). Second, by developing and using numerical methods for the solution of the lubrication equations, and third by comparing the high-order asymptotic results for the Trouton ratio, and the corresponding transformed formulas, with the exact analytical solution which is valid beyond the classic lubrication limit.

The rest of the paper is organized as follows. In § 2, we describe the flow geometry, the main dimensionless quantities, the governing equations and accompanied auxiliary conditions for the steady flow of an incompressible viscoelastic fluid in an axisymmetric non-uniform pipe. In § 3, we develop the general theoretical framework leading to general analytical results for the average pressure drop and the dimensionless elongational viscosity of the fluid (usually known as the Trouton ratio). In § 4 we derive a simplified set of governing equations, along with the analytical solutions for a couple of limiting cases. The method of developing high-order regular perturbation solutions in terms of the Deborah number for the lubrication equations is described in § 5. There, we report the most important new asymptotic formulas along with suitable convergence acceleration results for the average pressure-drop and the Trouton ratio. In § 6, we present the numerical methodology developed to solve the final lubrication equations, along with a comparison of the numerical results with the formulas derived in § 5. In § 7 we discuss an important finding of the analysis, namely a non-algebraic dependence of the solution for the Trouton ratio with respect to the Deborah number. Finally, in § 8 we provide our conclusions.

## 2. Problem formulation

The geometry and its main features are shown in Figure 1. A varying axisymmetric pipe is configured vertically, with its axis of symmetry aligned with the direction of gravity. The pipe consist of three segments; an entrance region with constant radius $h_0^*$, a varying region with radius $h_0^*$ at the inlet and radius $h_f^*$ at the outlet, and an exit region with constant radius $h_f^*$; the length of these regions are $\ell_{ent}^*$, $\ell^*$, and $\ell_{ex}^*$, respectively. We also assume a constant volumetric flow-rate, $Q^*$. For later convenience, and to facilitate the discussion and analysis, we define the aspect ratio of the varying region of the pipe, as well as one half of the average inlet velocity:

$$\varepsilon \equiv \frac{h_0^*}{\ell^*}, \quad u_c^* \equiv \frac{Q^*}{2\pi h_0^{*2}} \tag{2.1a,b}$$

For narrow and confined geometries such as that shown in Fig.1 the aspect ratio is small, $\varepsilon < 1$, a feature that will be exploited in § 4 to simplify the original governing equations. We consider a complex viscoelastic fluid (a polymeric material into a Newtonian solvent) with longest relaxation time $\lambda^*$, and we recognize two distinct time scales associated with the flow conditions for the process. The first is an inverse shear-rate $h_0^*/u_c^*$, and the second is an average residence time of the fluid in the pipe $\ell^*/u_c^*$. Based on these three characteristic times, we define two dimensionless groups of importance: the Weissenberg and Deborah numbers:

$$Wi \equiv \frac{\lambda^*}{h_0^*/u_c^*} = \frac{\lambda^* u_c^*}{h_0^*}, \quad De \equiv \frac{\lambda^*}{\ell^*/u_c^*} = \frac{\lambda^* u_c^*}{\ell^*} \qquad (2.2a,b)$$

from where we observe that $De = \varepsilon\, Wi$. Since the analysis performed here assumes a confined and narrow geometry, $0 < \varepsilon < 1$, and in order to be able to observe the effect of viscoelasticity, we demand that the Deborah number is finite when the aspect ratio goes to zero, which in turn implies that the Weissenberg number is large, i.e. $De = O(1)$ and $Wi = O(1/\varepsilon)$ as $\varepsilon \to 0$. Note that if we had assumed $Wi = O(1)$ as $\varepsilon \to 0$, then $De \to 0$ and either the relaxation time would be very small or the residence time would be very large. In the first case, the effect of viscoelasticity is negligible, while in the second its effect would be observed in a very small region near the entrance of the pipe. Hence, we proceed with $De = O(1)$ and $Wi = O(1/\varepsilon)$ as $\varepsilon \to 0$, and we keep in mind that a small value of $De$ means that the relaxation time of the polymeric material is small compared to the residence time (i.e., the flow is "slow"), while a large value of $De$ means that the polymer molecules do not have enough time to relax before their exit from the tube (i.e., the flow is "fast").

### 2.1. Governing equations

We consider the isothermal and incompressible steady flow of a viscoelastic fluid which consists of a Newtonian solvent with constant shear viscosity $\eta_s^*$ and a polymeric material with zero shear-rate viscosity $\eta_p^*$ and longest relaxation time $\lambda^*$. The fluid has constant mass density denoted by $\rho^*$. We use a cylindrical coordinate system $(z^*, y^*, \theta)$ to describe the flow field, where $z^*$ is the main flow direction which is aligned with the direction of gravity, $y^*$ is the radial direction, and $\theta$ is the azimuthal angle; $\mathbf{e}_\theta, \mathbf{e}_z$ and $\mathbf{e}_y$ are the unit vectors in the $\theta$, $z^*$, and $y^*$ directions, respectively. The origin of the coordinate system is placed on the center of the inlet cross-section with radius $h_0^*$ (see Figure 1). The wall of the pipe is described by the shape function $H^* = H^*(z^*) > 0$ for $z^* \in [0, \ell^*]$, i.e., $y^* = H^*(z^*)$. The shape function $H^*$ is considered fixed and known in advance; for $H^*(z^*) = h_0^*$ one gets a straight circular pipe. Also, we exclude variations of the flow field in the azimuthal angle which implies that the flow is axisymmetric and two-dimensional. The velocity vector in the flow domain is denoted by $\mathbf{u}^* = V^*(y^*,z^*)\mathbf{e}_y + U^*(y^*,z^*)\mathbf{e}_z$ and the total pressure is denoted by $p^* = p^*(y^*,z^*)^*$. The latter is induced by the flow and the imposed constant flow rate $Q^*$ at the inlet:

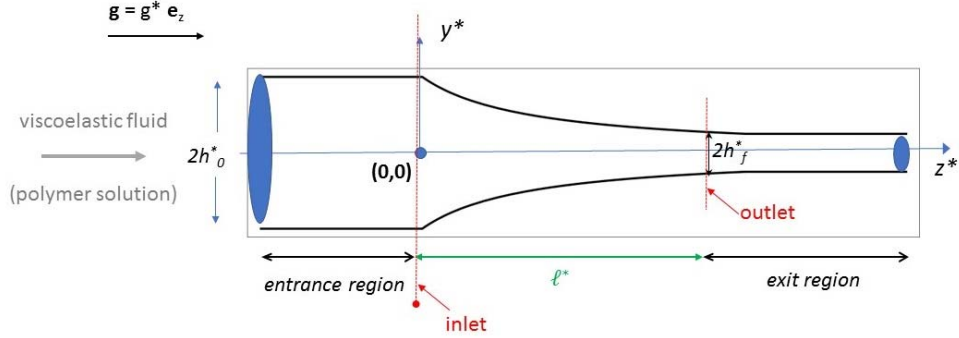

**Figure 1:** Cylindrical coordinate system ($y^*,z^*$) for an axisymmetric hyperbolic pipe with aspect ratio $\varepsilon=1/4$ and $\Lambda = h^*_0/h^*_f = 4$
The origin of the coordinate system is located at point (0,0)

$$Q^* = 2\pi \int_0^{h_0^*} U^*(y^*,0) y^* dy^* \qquad (2.3)$$

Using the unit tensor $\mathbf{I}$, the rate of deformation tensor $\dot{\boldsymbol{\gamma}}^* = \nabla^* \mathbf{u}^* + (\nabla^* \mathbf{u}^*)^T$ where

$$\nabla^* \equiv \mathbf{e}_y \frac{\partial}{\partial y^*} + \mathbf{e}_z \frac{\partial}{\partial z^*} + \mathbf{e}_\theta \frac{1}{y^*} \frac{\partial}{\partial \theta} \qquad (2.4)$$

is the gradient operator, the viscoelastic extra-stress tensor $\boldsymbol{\tau}^*$, the Reynolds stress tensor $\mathbf{u}^* \mathbf{u}^*$, and including the gravitational contribution in the vertical direction, $z^*$, we define the total momentum tensor:

$$\mathbf{T}^* := -p^* \mathbf{I} + \eta_s^* \dot{\boldsymbol{\gamma}}^* + \boldsymbol{\tau}^* - \rho^* \mathbf{u}^* \mathbf{u}^* + \rho^* g^* z^* \mathbf{e}_z \qquad (2.5)$$

In the absence of any other external forces and torques, the conservation equations that govern the flow in the pipe are the mass and momentum balances, respectively:

$$\nabla^* \cdot \mathbf{u}^* = 0, \quad \nabla^* \cdot \mathbf{T}^* = \mathbf{0} \qquad (2.6a,b)$$

Eqs (2.6a,b) are accompanied with a constitutive equation that models the response of the polymeric material to the flow deformation, namely with an equation for $\boldsymbol{\tau}^*$. Specifically, we consider the fundamental non-linear differential Oldroyd-B models, given by:

$$\boldsymbol{\tau}^* + \lambda^* \frac{\delta^* \boldsymbol{\tau}^*}{\delta t^*} = \eta_p^* \dot{\boldsymbol{\gamma}}^* \qquad (2.7)$$

where $\delta^*\boldsymbol{\tau}^*/\delta t^*$ represents the upper convected derivative of $\boldsymbol{\tau}^*$ at steady state:

$$\frac{\delta^*\boldsymbol{\tau}^*}{\delta t^*} = \mathbf{u}^* \cdot \nabla^*\boldsymbol{\tau}^* - \boldsymbol{\tau}^* \cdot \nabla^*\mathbf{v}^* - (\nabla^*\mathbf{v}^*)^T \cdot \boldsymbol{\tau}^* \quad (2.8)$$

The domain of definition of Eqs. (2.4)-(2.8) is

$$\Omega^* = \{(z^*, y^*, \theta) \mid 0 < z^* < \ell^*, 0 < y^* < H^*(z^*), 0 \leq \theta < 2\pi\} \quad (2.9)$$

The governing equations are solved with no-slip and no-penetration boundary conditions along the wall of the pipe

$$U^*(H^*(z^*), z^*) = V^*(H^*(z^*), z^*) = 0 \quad \text{at } 0 \leq z^* \leq \ell^* \quad (2.10)$$

Also, the equations are well-defined provided the conditions at $y^* = 0$:

$$V^*(0, z^*) = \left.\frac{\partial U^*}{\partial y^*}\right|_{y^*=0} = \tau_{yz}^*(0, z^*) = 0 \quad \text{at } 0 \leq z^* \leq \ell^* \quad (2.11a)$$

$$\tau_{yy}^*(0, z^*) = \tau_{\theta\theta}^*(0, z^*) = 0 \quad \text{at } 0 \leq z^* \leq \ell^* \quad (2.11b)$$

The integral constraint of mass at any distance from the inlet is also utilized:

$$Q^* = 2\pi \int_0^{H^*(z^*)} U^*(y^*, z^*) y^* dy^* = \text{constant at } 0 \leq z^* \leq \ell^* \quad (2.12)$$

Finally, a datum pressure, $p_{ref}^*$, is chosen at the wall of the outlet cross-section:

$$p_{ref}^* = p^*(H^*(\ell^*), \ell^*) - \rho^* g^* \ell^* \quad (2.13)$$

where the constant term $\rho^* g^* \ell^*$ has been subtracted from $p^*(H^*(\ell^*), \ell^*)$ for later convenience. Note that when the Newtonian solvent is absent ($\eta_s^* = 0$) the governing equations reduce to the Upper Convected Maxwell (UCM) model, or to the Oldroyd-B model when both the Newtonian solvent and the polymer molecules are present ($\eta_s^*, \eta_p^* > 0$; see below for more details).

## 2.2. Scaling and non-dimensionalization

Dimensionless variables are introduced based on the lubrication theory using the transformation $X = X^*/X_c^*$ where $X^* = z^*, y^*, H^*, U^*, V^*, P^*, \tau_{zz}^*, \tau_{yz}^*, \tau_{yy}^*, \tau_{\theta\theta}^*$ and $X_c^*$ is the relevant characteristic scale for $X^*$. The characteristic scales are defined with the aid of the aspect ratio of the pipe and one half of the cross-sectionally average velocity at the inlet of the pipe defined in Eq. (2.1a,b). The $z^*$-coordinate is scaled by $\ell^*$ and the $y^*$-coordinate and the shape function $H^*$ by $h_0^*$. The characteristic scale for the main velocity component $U^*$ is $u_c^*$, and from the continuity equation one finds that the characteristic scale for the vertical velocity component $V^*$ is $\varepsilon u_c^*$. With the aid of the momentum balance along the main flow direction, $z^*$, the characteristic scale for the pressure difference $p^* - p_{ref}^*$ is $(\eta_s^* + \eta_p^*) u_c^* \ell^* / h_0^{*2}$ (Tavakol *et al.* 2017, Housiadas & Tsagaris 2022, 2023; Housiadas & Beris 2023). For the viscoelastic extra-stress components, $\tau_{zz}^*, \tau_{yz}^*, \tau_{yy}^*$ and $\tau_{\theta\theta}^*$, consistency of the governing equations for a negligible relaxation time of the polymer ($\lambda^* \to 0$) and a long pipe ($h_0^* \ll \ell^*$) implies that different scales for the components of $\boldsymbol{\tau}^*$ must be used. Careful examination of the momentum balance and the individual components of the constitutive equation shows that the characteristic scales for $\tau_{yz}^*$, $\tau_{zz}^*$, $\tau_{yy}^*$ and $\tau_{\theta\theta}^*$ are $\eta_p^* u_c^* / h_0^*$, $\eta_p^* u_c^* \ell^* / h_0^{*2}$, $\eta_p^* u_c^* / \ell^*$ and $\eta_p^* u_c^* / \ell^*$, respectively. Based on these scales, the velocity vector and the Reynolds stress tensor are given as, respectively

$$\mathbf{u} = U \mathbf{e}_z + \varepsilon V \mathbf{e}_y \tag{2.19}$$

$$\mathbf{uu} = \mathbf{e}_z \mathbf{e}_z U^2 + \mathbf{e}_z \mathbf{e}_y \varepsilon U V + \mathbf{e}_y \mathbf{e}_z \varepsilon U V + \mathbf{e}_y \mathbf{e}_y \varepsilon^2 V^2 \tag{2.20}$$

Similarly, with the aid of the gradient operator

$$\nabla \equiv h_0^* \nabla^* = \mathbf{e}_y \frac{\partial}{\partial y} + \varepsilon \mathbf{e}_z \frac{\partial}{\partial z} + \mathbf{e}_\theta \frac{1}{y} \frac{\partial}{\partial \theta} \tag{2.21}$$

we find that the rate of deformation tensor, scaled by $u_c^* / h_0^*$, is

$$\dot{\boldsymbol{\gamma}} = \mathbf{e}_z \mathbf{e}_z \left( 2\varepsilon \frac{\partial U}{\partial z} \right) + \mathbf{e}_z \mathbf{e}_y \left( \frac{\partial U}{\partial y} + \varepsilon^2 \frac{\partial V}{\partial z} \right) + \mathbf{e}_y \mathbf{e}_z \left( \frac{\partial U}{\partial y} + \varepsilon^2 \frac{\partial V}{\partial z} \right) + \mathbf{e}_y \mathbf{e}_y \left( 2\varepsilon \frac{\partial V}{\partial y} \right) + \mathbf{e}_\theta \mathbf{e}_\theta \left( 2\varepsilon \frac{V}{y} \right) \tag{2.22}$$

Also, the dimensionless polymer extra-stress tensor, scaled by $\eta_p^* u_c^* / h_0^*$, is:

$$\boldsymbol{\tau} = \mathbf{e}_z \mathbf{e}_z \tau_{zz} / \varepsilon + \mathbf{e}_z \mathbf{e}_y \tau_{yz} + \mathbf{e}_y \mathbf{e}_z \tau_{yz} + \mathbf{e}_y \mathbf{e}_y \varepsilon \tau_{yy} + \mathbf{e}_\theta \mathbf{e}_\theta \varepsilon \tau_{\theta\theta} \tag{2.23}$$

Finally, the total momentum tensor, scaled by $(\eta_s^* + \eta_p^*) u_c^* \ell^* / h_0^{*2}$, is given by:

$$\mathbf{T} = -p \mathbf{I} + \varepsilon (1 - \eta) \dot{\boldsymbol{\gamma}} + \varepsilon \eta \boldsymbol{\tau} - \mathrm{Re}_m \mathbf{uu} + St\, z\, \mathbf{e}_z \tag{2.24}$$

In Eq. (2.24), the modified Reynolds number, $\mathrm{Re}_m$, the polymer viscosity ratio, $\eta$, and the Stokes number, $St$, appear:

$$\mathrm{Re}_m \equiv \frac{\rho^* u_c^* h_0^{*2}}{(\eta_s^* + \eta_p^*) \ell^*}, \quad \eta \equiv \frac{\eta_p^*}{\eta_s^* + \eta_p^*}, \quad St \equiv \frac{\rho^* g^* h_0^{*2}}{(\eta_s^* + \eta_p^*) u_c^*} \tag{2.25}$$

Defining the modified pressure $P := p - St\, z$, the dimensionless form of the balance equations in scalar form is:

$$\frac{\partial U}{\partial z}+\frac{\partial V}{\partial y}+\frac{V}{y}=0 \tag{2.26}$$

$$\text{Re}_m \frac{DU}{Dt} = -\frac{\partial P}{\partial z} + (1-\eta)\left(\frac{\partial^2 U}{\partial y^2} + \frac{1}{y}\frac{\partial U}{\partial y} + \varepsilon^2 \frac{\partial^2 U}{\partial z^2}\right) + \eta\left(\frac{\partial \tau_{yz}}{\partial y} + \frac{\tau_{yz}}{y} + \frac{\partial \tau_{zz}}{\partial z}\right) \tag{2.27}$$

$$\varepsilon^2 \text{Re}_m \frac{DV}{Dt} = -\frac{\partial P}{\partial y} + (1-\eta)\varepsilon^2\left(\frac{\partial^2 V}{\partial y^2} + \frac{1}{y}\frac{\partial V}{\partial y} - \frac{V}{y^2} + \varepsilon^2 \frac{\partial^2 V}{\partial z^2}\right) + \eta\varepsilon^2\left(\frac{\partial \tau_{yy}}{\partial y} + \frac{\partial \tau_{yz}}{\partial z} + \frac{\tau_{yy}-\tau_{\theta\theta}}{y}\right) \tag{2.28}$$

where $D/Dt$ is the material derivative defined at steady state as

$$\frac{D}{Dt} \equiv U\frac{\partial}{\partial z} + V\frac{\partial}{\partial y} \tag{2.29}$$

For $\eta=0$, i.e., in absence of the polymeric molecules, Eqs. (2.26)-(2.29) reduce to the well-known dimensionless Navier-Stokes equations for an incompressible Newtonian fluid. Notice that for $\eta=1$ there is no solvent contribution, i.e., the fluid is a pure polymeric material, while for $0<\eta<1$ the fluid consists of a Newtonian solvent and a polymeric material.

The UCM/Oldroyd-B models in scalar form are given as:

$$\tau_{zz} + De\left(\frac{D\tau_{zz}}{Dt} - 2\tau_{zz}\frac{\partial U}{\partial z} - 2\tau_{yz}\frac{\partial U}{\partial y}\right) = 2\varepsilon^2 \frac{\partial U}{\partial z} \tag{2.30}$$

$$\tau_{yy} + De\left(\frac{D\tau_{yy}}{Dt} - 2\tau_{yz}\frac{\partial V}{\partial z} - 2\tau_{yy}\frac{\partial V}{\partial y}\right) = 2\frac{\partial V}{\partial y} \tag{2.31}$$

$$\tau_{\theta\theta} + De\left(\frac{D\tau_{\theta\theta}}{Dt} - 2\tau_{\theta\theta}\frac{V}{y}\right) = 2\frac{V}{y} \tag{2.32}$$

$$\tau_{yz} + De\left(\frac{D\tau_{yz}}{Dt} + \tau_{yz}\frac{V}{y} - \tau_{zz}\frac{\partial V}{\partial z} - \tau_{yy}\frac{\partial U}{\partial y}\right) = \frac{\partial U}{\partial y} + \varepsilon^2 \frac{\partial V}{\partial z} \tag{2.33}$$

The dimensionless domain of definition of Eqs. (2.26)-(2.33) is:

$$\Omega = \{(z,y,\theta) \mid 0<z<1, 0<y<H(z), 0\leq\theta<2\pi\} \tag{2.34}$$

Various cases can be considered for the shape function but here we focus on the hyperbolic pipe:

$$H(z) = \frac{1}{\sqrt{(\Lambda^2-1)z+1}}, \quad 0\leq z\leq 1 \tag{2.35}$$

where $\Lambda = h_0^*/h_f^*$. In the entrance region ($z\leq 0$) $H(z)=1$, while in the exit region ($z\geq 1$), $H(z)=1/\Lambda$. Thus, $H=H(z)$ is a piecewise smooth function in the regions $(-\ell_{en}^*/\ell^*,0)$, $(0,1)$ and $(0,\ell_{ex}^*/\ell^*)$. It is continuous in the whole domain $(-\ell_{en}^*/\ell^*,\ell_{ex}^*/\ell^*)$, but not differentiable at $z=0$ and $z=1$. For $\Lambda >1$, Eq. (2.35) describes a converging pipe, while for $0<\Lambda <1$ it describes an expanding pipe; in the former case Λ will be referred to as the contraction ratio.

The auxiliary (boundary, symmetry, and integral) dimensionless conditions are:

$$V = U = 0 \quad \text{at} \quad y = H(z), 0 \leq z \leq 1 \tag{2.36a}$$

$$V = \frac{\partial U}{\partial y} = \tau_{yz} = 0 \quad \text{at} \quad y = 0, 0 \leq z \leq 1 \tag{2.36b}$$

$$\tau_{yy} = \tau_{\theta\theta} \quad \text{at} \quad y = 0, 0 \leq z \leq 1 \tag{2.36c}$$

$$\int_0^{H(z)} U(y,z)\, y\, dy = 1, \ 0 \leq z \leq 1 \tag{2.36d}$$

$$P(H(1),1) = 0 \tag{2.36e}$$

Equations (2.36b,c) are needed so that Eqs (2.26)-(2.28) are well-defined at *y*=0, while equation (2.36d) results from the integration of Eq. (2.26) along the *y*-direction, between the limits $y=0$ and $y=H(z)$, applying the no-penetration boundary condition and using the dimensionless flow rate at the inlet of the pipe.

## 2.3. The conformation tensor

As mentioned above, the constitutive models for the polymeric molecules are frequently provided in terms of the conformation tensor $\mathbf{C}$ (dimensionless). For the Oldroyd-B model, and its limiting form, the UCM model, the conformation tensor is related to the extra-stress tensor (Bird *et al.* 1987; Beris & Edwards 1994) as follows:

$$\boldsymbol{\tau}^* = \frac{\eta_p^*}{\lambda^*}(\mathbf{C}-\mathbf{I}) \tag{2.37}$$

For a Newtonian fluid, i.e. for $\lambda^* \to 0$, the conformation tensor reduces to the identity tensor, $\mathbf{C} \to \mathbf{I}$, and the extra-stress tensor to the rate-of-deformation tensor, $\boldsymbol{\tau}^* \to \dot{\boldsymbol{\gamma}}^*$; for no-flow $\mathbf{C} \to \mathbf{I}$ and $\boldsymbol{\tau}^* \to \mathbf{0}$. An important property of the real, second-order and symmetric tensor $\mathbf{C}$ is that is positive definite, and therefore its eigenvalues and its determinant, are strictly positive.

Based on the dimensionless form of the extra-stress tensor, $\boldsymbol{\tau}$, the conformation tensor, $\mathbf{C}$, is given through $\boldsymbol{\tau} = (\mathbf{C}-\mathbf{I})/Wi = \varepsilon(\mathbf{C}-\mathbf{I})/De$ where:

$$\mathbf{C} = \mathbf{e}_z\mathbf{e}_z \underbrace{(\varepsilon^2 C_{zz})}_{c_{zz}}/\varepsilon^2 + \mathbf{e}_z\mathbf{e}_y \underbrace{(\varepsilon C_{yz})}_{c_{yz}}/\varepsilon + \mathbf{e}_y\mathbf{e}_z \underbrace{(\varepsilon C_{yz})}_{c_{yz}}/\varepsilon + \mathbf{e}_y\mathbf{e}_y \underbrace{C_{yy}}_{c_{yy}} + \mathbf{e}_\theta\mathbf{e}_\theta \underbrace{C_{\theta\theta}}_{c_{\theta\theta}} \tag{2.38a}$$

or

$$\mathbf{C} = \mathbf{e}_z\mathbf{e}_z c_{zz}/\varepsilon^2 + \mathbf{e}_z\mathbf{e}_y c_{yz}/\varepsilon + \mathbf{e}_y\mathbf{e}_z c_{yz}/\varepsilon + \mathbf{e}_y\mathbf{e}_y c_{yy} + \mathbf{e}_\theta\mathbf{e}_\theta c_{\theta\theta} \qquad (2.38b)$$

From Eqs (2.23) and (2.38b), we find that the components of $\boldsymbol{\tau}$ and the rescaled components of $\mathbf{C}$ are connected as:

$$\tau_{zz} = \frac{c_{zz} - \varepsilon^2}{De}, \quad \tau_{yz} = \frac{c_{yz}}{De}, \quad \tau_{yy} = \frac{c_{yy} - 1}{De}, \quad \tau_{\theta\theta} = \frac{c_{\theta\theta} - 1}{De} \qquad (2.39)$$

One is easy to confirm that the Oldroyd-B model can be equivalently, and consistency expressed in terms of $\tau_{zz}, \tau_{yz}, \tau_{yy}$ and $\tau_{\theta\theta}$ or $c_{zz}, c_{yz}, c_{yy}$ and $c_{\theta\theta}$; the same expressions for $\tau_{zz}$, $\tau_{yz}$ and $\tau_{yy}$ have also been derived by Boyko & Stone (2022), Ahmed & Biancofiore (2021, 2023) and Housiadas & Beris (2023) for the viscoelastic lubrication flow in a hyperbolic channel. Finally, we mention that an unstretched initial condition for the polymer molecules implies that

$$c_{zz} = \varepsilon^2, \quad c_{yy} = c_{\theta\theta} = 1, \quad c_{yz} = 0 \qquad (2.40a)$$

or, in terms of the components of the original conformation tensor as follows:

$$C_{zz} = C_{yy} = C_{\theta\theta} = 1, \quad C_{yz} = 0 \qquad (2.40b)$$

## 3. General analytical results

The main quantities of interest for this type of flow are the average pressure-drop required to maintain the constant flowrate through the pipe, the first normal stress difference and the Trouton ratio (or dimensionless elongational viscosity) of the fluid, as well as the viscoelastic extra-stresses at the wall.

### 3.1. Pressure-drop

The average pressure-drop is defined as:

$$\Delta\Pi^* := 2\int_0^{H^*(0)} P^*(y^*,0) y^* dy^* - 2\int_0^{H^*(1)} P^*(y^*,1) y^* dy^* = -2\int_0^1 \frac{d}{dz^*}\left(\int_0^{H^*(z^*)} P(y^*,z) y^* dy^*\right) dz^* \qquad (3.1)$$

Using the characteristic scales reported in § 2.3, we find $\Delta\Pi \equiv \Delta\Pi^*/\Delta\Pi_c^*$:

$$\Delta\Pi = 2\int_0^{H(0)} P(y,0) y\,dy - 2\int_0^{H(1)} P(y,1) y\,dy = -2\int_0^1 \frac{d}{dz}\left(\int_0^{H(z)} P(y,z) y\,dy\right) dz \qquad (3.2)$$

where $\Delta\Pi_c^* = (\eta_s^* + \eta_t^*)\ell^* Q^*/(2h_0^{*2})$. Applying Leibniz's rule for integrals in Eq. (3.2), and splitting the resulting expression, we find:

$$\Delta\Pi = -2\int_0^1\left\{\int_0^H (\frac{\partial P}{\partial z} y\, dy) dz\right\} - \int_0^1 2H'H\, P(H,z) dz \quad (3.3)$$

The second term on the right-hand-side of Eq. (3.3) is simplified recalling that $H = H(z)$, using $2H'H = (H^2 - 1)'$ and performing integration by parts:

$$\int_0^1 2H'H\, P(H,z) dz = \left[(H^2(z)-1)P(H(z),z)\right]_0^1 - \int_0^1 (H^2-1)\frac{d}{dz}P(H,z) dz \quad (3.4)$$

Since $H(0)=1$ (due to dimensionalization) and $P(H(1),1)=0$ (due to the reference pressure), the first term on the right-hand-side of Eq. (3.4) vanishes. Thus, Eqs (3.3) and (3.4) yield:

$$\Delta\Pi = -2\int_0^1\left(\int_0^H \frac{\partial P}{\partial z} y\, dy\right) dz + \int_0^1 (H^2-1)\frac{d}{dz}P(H,z) dz \quad (3.5)$$

Using the chain rule to find the derivative of $P(H(z),z)$ with respect to z, we find the general formula for the average pressure drop in the varying region of the pipe:

$$\Delta\Pi = -2\int_0^1\left(\int_0^H (\frac{\partial P}{\partial z} y\, dy)\right) dz + \int_0^1 (H^2-1)\left(\left.\frac{\partial P}{\partial z}\right|_{y=H} + H'\left.\frac{\partial P}{\partial y}\right|_{y=H}\right) dz \quad (3.6)$$

All quantities in Eq. (3.6) can be obtained directly from the non-trivial components of the momentum balance, i.e., Eqs. (2.27)-(2.28), which can also be used to reveal the individual contributions to the average pressure drop.

### 3.2. First normal stress difference

The extensional character of the flow is investigated starting from the symmetry conditions at *y*=0, given by Eq. (2.36b), along with continuity equation evaluated at the axis of symmetry. In this case, we find:

$$\lim_{y\to 0}\frac{V}{y} = \left.\frac{\partial V}{\partial y}\right|_{y=0} = -\frac{1}{2}\left.\frac{\partial U}{\partial z}\right|_{y=0} \quad (3.7)$$

Due to Eq. (3.7), the rate-of-deformation tensor is given by:

$$\left.\dot{\gamma}\right|_{y=0} = 2\varepsilon\left.\frac{\partial U}{\partial z}\right|_{y=0}\left(\mathbf{e}_z\mathbf{e}_z - \frac{1}{2}\mathbf{e}_y\mathbf{e}_y - \frac{1}{2}\mathbf{e}_\theta\mathbf{e}_\theta\right) \quad (3.8)$$

Thus, the flow at the axis of symmetry is a pure uniaxial extensional flow and the quantity $2\varepsilon\,\partial U/\partial z|_{y=0}$ is a dimensionless rate of extension $\dot{E}$. We emphasize, however, that the flow in the hyperbolic pipe is a mixed flow with both extensional and shear characteristics. It is purely extensional in character along the axis of symmetry of the pipe only, while at the wall the shear characteristics dominate. When the rate-of-deformation tensor is given in the form of Eq. (3.8), the dimensionless extensional viscosity of the fluid can be determined merely from the first total normal stress difference $N_1 := (T_{zz} - T_{yy})/\varepsilon^2$ where $\varepsilon^2$ in the denominator has been introduced due to scaling. Following the definition of the total stress tensor given in Eq. (2.24), the rate of deformation tensor given in (2.22), and the viscoelastic extra-stress tensor given in Eq. (2.23), we find:

$$N_1 = 3(1-\eta)\frac{\partial U}{\partial z} + \eta\left(\frac{\tau_{zz}}{\varepsilon^2} - \tau_{yy}\right) \text{ at } y=0 \tag{3.9a}$$

For a Newtonian fluid, i.e., for $De = 0$, Eqs. (2.30) and (2.31) with the aid of Eq. (3.7) give $\tau_{zz,N} = 2\varepsilon^2\, \partial U/\partial z\big|_{y=0}$ and $\tau_{yy,N} = -\partial U/\partial z\big|_{y=0}$, and therefore Eq. (3.9a) reduces to $N_{1,N} = 3\,\partial U/\partial z\big|_{y=0}$. Eq. (3.9a) can also be expressed in terms of the rescaled components of the conformation tensor (see Eq. (2.38b)) as follows:

$$N_1 = 3(1-\eta)\frac{\partial U}{\partial z} + \frac{\eta}{De}\left(\frac{c_{zz}}{\varepsilon^2} - c_{yy}\right) \text{ at } y=0 \tag{3.9b}$$

or in terms of the original components of the conformation tensor:

$$N_1 = 3(1-\eta)\frac{\partial U}{\partial z} + \frac{\eta}{De}\left(C_{zz} - C_{yy}\right) \text{ at } y=0 \tag{3.9c}$$

Considering the constraints on the axis of symmetry (see Eq. (2.36b,c)), the $zz-$, $yy-$ and $\theta\theta-$ components of the Oldroyd-B model can be solved analytically. Denoting with $u(z) \equiv U(y=0,z)$ and $u'(z) \equiv \partial U/\partial z\big|_{y=0}$, and taking into account Eqs. (2.39) and (3.7), Eqs. (2.10)-(2.12) are given in terms of the rescaled components of the conformation tensor (see Eq. (2.38a)):

$$c_{zz} + De\left(u\, c'_{zz} - 2c_{zz}u'\right) = \varepsilon^2 \tag{3.10}$$

$$c_{yy} + De\left(u\, c'_{yy} + c_{yy}u'\right) = 1 \tag{3.11}$$

$$c_{\theta\theta} + De\left(u\, c'_{\theta\theta} + c_{\theta\theta}u'\right) = 1 \tag{3.12}$$

The exact analytical solution of Eqs (3.10)-(3.12) is:

$$c_{zz}(0,z) = u^2(z)\varphi(z)\left\{\frac{c_{zz}(0,0)}{u^2(0)} + \frac{\varepsilon^2}{De}\int_0^z \frac{dx}{\varphi(x)u^3(x)}\right\} \tag{3.13}$$

$$c_{yy}(0,z) = \frac{\varphi(z)}{u(z)}\left\{u(0)c_{yy}(0,0) + \frac{1}{De}\int_0^z \frac{dx}{\varphi(x)}\right\} \tag{3.14}$$

$$c_{\theta\theta}(0,z) = \frac{\varphi(z)}{u(z)}\left\{u(0)c_{\theta\theta}(0,0) + \frac{1}{De}\int_0^z \frac{dx}{\varphi(x)}\right\} \quad (3.15)$$

where

$$\varphi(z) := \exp\left(-\frac{1}{De}\int_0^z \frac{ds}{u(s)}\right) \quad (3.16)$$

From Eq. (2.39) and Eqs. (3.13)-(3.16), the normal extra-stress components are trivially found as:

$$\tau_{zz}(0,z) = \frac{c_{zz}(0,z) - \varepsilon^2}{De}, \quad \tau_{yy}(0,z) = \frac{c_{yy}(0,z) - 1}{De}, \quad \tau_{\theta\theta}(0,z) = \frac{c_{\theta\theta}(0,z) - 1}{De} \quad (3.17)$$

Moreover, since $C_{zz} \equiv c_{zz}/\varepsilon^2$, $C_{yy} \equiv c_{yy}$ and $C_{\theta\theta} \equiv c_{\theta\theta}$, the components of the original conformation tensor can easily be extracted too; worth mentioning is that the solution for $C_{zz}$, $C_{yy}$ and $C_{\theta\theta}$ at $y = 0$ depends on the aspect ratio only implicitly through the velocity profile $u = u(z)$. Eqs. (3.13)-(3.16) also show that the spatial evolution of $C_{zz}$, $C_{yy}$ and $C_{\theta\theta}$ at $y = 0$ depends on the initial conditions at the inlet of the pipe (as it should be expected) as well as on the fluid velocity at $y = 0$, $u = u(z)$. Noticing that $\int_0^z \frac{ds}{u(s)} > 0$ it is clear that $\varphi = \varphi(z)$ is a continuous and positive function with $\varphi(0) = 1$ which decreases exponentially in the range $0 < \varphi(z) \leq 1$.

Finally, by plugging Eqs (3.13)-(3.17) in Eq. (3.9b), we find:

$$N_1 = 3(1-\eta)\,u'(z) + \frac{\eta}{De^2}\varphi(z)\left(u^2(z)\int_0^z \frac{dx}{\varphi(x)u^3(x)} - \frac{1}{u(z)}\int_0^z \frac{dx}{\varphi(x)}\right) +$$
$$+ \frac{\eta}{De}\varphi(z)\left(\frac{u^2(z)c_{zz}(0,0)}{u^2(0)\varepsilon^2} - \frac{u(0)}{u(z)}c_{yy}(0,0)\right) \quad (3.18a)$$

Eq. (3.18a) is the general formula for the evaluation of the dimensionless first normal stress difference based on the flow field at the axis of symmetry of the pipe. Restoring the original components of the conformation tensor, $c_{zz} \equiv \varepsilon^2 C_{zz}$ and $c_{yy} \equiv C_{yy}$, and assuming that the polymer molecules are unstretched at the inlet of the varying region of the pipe, namely for $c_{zz}(0,0) = \varepsilon^2$ and $c_{yy}(0,0) = 1$ (see Eq. (2.40a)), or equivalently, $C_{zz}(0,0) = C_{yy}(0,0) = 1$ (see Eq. (2.40b)), gives:

$$N_1 = 3(1-\eta)\,u'(z) + \frac{\eta\varphi(z)}{De}\left(\frac{1}{De}\left(u^2(z)\int_0^z \frac{dx}{\varphi(x)u^3(x)} - \frac{1}{u(z)}\int_0^z \frac{dx}{\varphi(x)}\right) + \frac{u^2(z)}{u^2(0)} - \frac{u(0)}{u(z)}\right) \quad (3.18b)$$

We emphasize that Eqs. (3.18a) and (3.18b) are exact, namely no approximations at all have been made for their derivation, as well as that Eq. (3.18b) does not depend directly on the aspect ratio; there is an indirect dependence of $N_1$ on $\varepsilon^2$ through $u$ though.

### 3.3. Viscoelastic extra-stresses at the wall

Considering the boundary conditions for the velocity field at the wall eliminates the material derivative of the polymeric extra stress(es) along the wall, namely $Df/Dt = 0$ for $f = \tau_{zz}, \tau_{yz}, \tau_{yy}, \tau_{\theta\theta}$ at $y = H(z)$. One can also confirm that along the wall, all velocity gradients can be expressed in terms of the wall shear rate $\dot{\gamma} \equiv \partial U / \partial y|_{y=H}$ only. This can be shown starting from the no-slip and no-penetration conditions (Eq. (2.36a)), differentiating with respect to $z$ and using the continuity equation (Eq. (2.26)):

$$\frac{\partial U}{\partial z} = -H'\dot{\gamma}, \quad \frac{\partial V}{\partial y} = H'\dot{\gamma}, \quad \frac{\partial V}{\partial z} = -H'^2\dot{\gamma} \quad \text{at } y = H \tag{3.19}$$

Using Eq. (3.19), the exact analytical solution for the constitutive model, Eqs (2.30)-(2.33), is:

$$\begin{aligned}
\tau_{zz}(H,z) &= 2De\,\dot{\gamma}^2 + 2\varepsilon^2\dot{\gamma}H'(-1+De\,H'\dot{\gamma}) \\
\tau_{yy}(H,z) &= 2H'\dot{\gamma}(1+De\,H'\dot{\gamma}) + 2\varepsilon^2 De\,\dot{\gamma}^2 H'^4 \\
\tau_{\theta\theta}(H,z) &= 0 \\
\tau_{yz}(H,z) &= \dot{\gamma}(1+2De\,H'\dot{\gamma}) + \varepsilon^2\dot{\gamma}H'^2(-1+2De\,\dot{\gamma}H')
\end{aligned} \tag{3.20}$$

Eq. (3.20) is well-defined for any value of $\dot{\gamma}$ and $De \geq 0$. Moreover, based on the positive definiteness criterion of the confirmation tensor $\mathbf{C}$, its diagonal elements and its determinant, $\det(\mathbf{C}) = C_{\theta\theta}(C_{yy}C_{zz} - C_{yz}^2) = c_{\theta\theta}(c_{yy}c_{zz} - c_{yz}^2)/\varepsilon^2$, must be positive. Indeed, by using Eqs. (2.39) and (3.20) we have confirmed that the diagonal elements of $\mathbf{C}$ and its determinant

$$\det(\mathbf{C}) = \frac{\varepsilon^2 + (De\,\dot{\gamma})^2(1+\varepsilon^2H'^2)^2}{\varepsilon^2} = 1 + (Wi\,\dot{\gamma})^2(1+\varepsilon^2H'^2)^2 \tag{3.21}$$

are strictly positive for any non-negative value of $De$, or $Wi$, and any geometry. Finally, for later use, we report the tangential viscoelastic extra-stress at the wall, $\tau_{nt}$, which can be found by calculating the quantity $(\boldsymbol{\tau}\cdot\mathbf{n})\cdot\mathbf{t}|_{y=H} \equiv \tau_{nt}$, where $\mathbf{n} = (\varepsilon H'\mathbf{e}_z - \mathbf{e}_y)/\sqrt{1+\varepsilon^2 H'^2}$ and $\mathbf{t} = (\mathbf{e}_z + \varepsilon H'\mathbf{e}_y)/\sqrt{1+\varepsilon^2 H'^2}$ are the normal and tangential unit vectors at the wall, respectively, and using Eq. (3.20). In this case, we find:

$$\tau_{nt} = \frac{\tau_{yz} - H'\tau_{zz}}{1+\varepsilon^2 H'^2} = \dot{\gamma} \quad \text{at } y = H \tag{3.22}$$

## 4. Lubrication approximation

The evaluation of the average pressure drop, Eq. (3.6), the fist normal stress difference, Eq. (3.18b), and the viscoelastic extra-stresses at the wall, Eq. (3.20), require the velocity field in the flow domain $\Omega$. We won't proceed by solving the full non-linear governing equations reported in § 2, but we will attempt to find good approximations of the field variables. This can be achieved by considering pipes with small aspect ratio, i.e., $\varepsilon < 1$. From an asymptotic point of view and based on the magnitude of $\varepsilon$, all terms in Eqs. (2.26)-(2.28) and the constitutive equation, Eqs. (2.30)-(2.33), multiplied with $\varepsilon^2$ or $\varepsilon^4$ are much smaller than the other terms and can be ignored as a first approximation. Therefore, the solution for each dependent field variable $X$ can be expressed formally as a perturbation power series in terms of $\varepsilon^2$, $X(y,z) \approx \sum_{i \geq 0} X_{(2i)}(y,z)\varepsilon^{2i}$ as $\varepsilon^2 \to 0$. Dropping the zero subscript in parenthesis throughout the paper for simplicity, unless otherwise noted, the leading-order balance equations are:

$$\frac{\partial U}{\partial z} + \frac{\partial V}{\partial y} + \frac{V}{y} = 0 \tag{4.1}$$

$$\mathrm{Re}_m \frac{DU}{Dt} = -\frac{\partial P}{\partial z} + (1-\eta)\left(\frac{\partial^2 U}{\partial y^2} + \frac{1}{y}\frac{\partial U}{\partial y}\right) + \eta\left(\frac{\partial \tau_{zz}}{\partial z} + \frac{\partial \tau_{yz}}{\partial y} + \frac{\tau_{yz}}{y}\right) \tag{4.2}$$

$$0 = -\frac{\partial P}{\partial y} \tag{4.3}$$

$$\tau_{zz} + De\left(\frac{D\tau_{zz}}{Dt} - 2\tau_{zz}\frac{\partial U}{\partial z} - 2\tau_{yz}\frac{\partial U}{\partial y}\right) = 0 \tag{4.4a}$$

$$\tau_{yy} + De\left(\frac{D\tau_{yy}}{Dt} - 2\tau_{yz}\frac{\partial V}{\partial z} - 2\tau_{yy}\frac{\partial V}{\partial y}\right) = 2\frac{\partial V}{\partial y} \tag{4.4b}$$

$$\tau_{\theta\theta} + De\left(\frac{D\tau_{\theta\theta}}{Dt} - 2\tau_{\theta\theta}\frac{V}{y}\right) = 2\frac{V}{y} \tag{4.4c}$$

$$\tau_{yz} + De\left(\frac{D\tau_{yz}}{Dt} + \tau_{yz}\frac{V}{y} - \tau_{zz}\frac{\partial V}{\partial z} - \tau_{yy}\frac{\partial U}{\partial y}\right) = \frac{\partial U}{\partial y} \tag{4.4d}$$

where the convective derivative $D/Dt$ is given by Eq. (2.29). The same type of lubrication analysis has been performed before for the planar case by Boyko & Stone (2022), Ahmed & Biancofiore (2021, 2023) and Housiadas & Beris (2023). However, as it will be become clear later, the analysis for the cylindrical pipe requires special attention because from the mathematical point of view, the axis of symmetry of the pipe is singular, namely the governing equations are not strictly defined on *y*=0, but they should be considered only as the limit as *y* goes to zero. This implies that the conditions (2.36b) and (2.36c) must hold.

Eq. (4.3) shows that $P = P(z)$ only and thus $\partial P / \partial z \equiv P'(z)$ is an unknown function that must be determined; this is a major feature of the classic lubrication theory at the limit of a vanishing small aspect ratio. It is also interesting that the corresponding equation for $\tau_{\theta\theta}$, Eq. (4.4c), is decoupled from the remaining equations. Thus, one can solve Eqs (4.1), (4.2) and (4.4a,b,d) to find the velocity components $U$ and $V$, the pressure gradient $P'(z)$ and the viscoelastic extra-stresses $\tau_{zz}, \tau_{yy}$ and $\tau_{yz}$, and then Eq. (4.4d) to find $\tau_{\theta\theta}$ for completeness. This decoupling is consequence of the lubrication approximation which neglects $\tau_{\theta\theta}$ in the momentum balance, and the axisymmetry of the flow field.

We proceed by reporting the analytical solution of the lubrication equations at the entrance region of the pipe, the analytical solution of the lubrication equations in the varying region of the pipe for a Newtonian fluid at creeping flow conditions, and by providing information about the discontinuity of the viscoelastic extra-stress when the polymeric fluid exits the entrance region and enters the varying region. Most importantly, starting from the general expressions derived in § 3, we derive the simplified formulas for the average pressure drop, the first normal stress difference, and the Trouton ratio at the classic lubrication limit.

### 4.1. Solution at the entrance region

The pipe at the entrance region has a constant radius, i.e., $H(z) = 1$ for $z \leq 0$. In this case, Eqs. (4.1)-(4.4a-d) can easily be solved analytically for all the field variables:

$$U_{en} = 4(1 - y^2), \quad V_{en} = 0, \quad P'_{en} = -16$$
$$\tau_{zz,en} = 128 De\, y^2, \quad \tau_{yz,en} = -8y, \quad \tau_{yy,en} = \tau_{\theta\theta,en} = 0 \qquad (4.5)$$

where the subscript "*en*" is used to denote the entrance region.

### 4.2. Newtonian solution for creeping flow

For a Newtonian fluid (*De*=0) and creeping flow conditions ($\text{Re}_m = 0$) Eqs. (4.1)-(4.3) can be solved analytically too (Housiadas & Tsangaris 2022). The solution is:

$$U_N = \frac{4}{H^2}\left(1 - \frac{y^2}{H^2}\right), \quad V_N = \frac{H'}{H} y U_N, \quad P'_N = -\frac{16}{H^4} \qquad (4.6)$$

where a subscript "*N*" throughout the paper denotes the solution for the Newtonian fluid. Using Eq. (4.6), one can find the polymer extra-stresses through Eqs. (4.4a-d):

$$\tau_{zz,N} = 0, \quad \tau_{yz,N} = \frac{\partial U_N}{\partial y} = -\frac{8y}{H^4},$$

$$\tau_{yy,N} = 2\frac{\partial V_N}{\partial y} = \frac{8H'}{H^3}\left(1 - 3\frac{y^2}{H^2}\right), \quad \tau_{\theta\theta,N} = 2\frac{V_N}{y} = \frac{8H'}{H^3}\left(1 - \frac{y^2}{H^2}\right) \quad (4.7)$$

### 4.3. Extra-stresses discontinuity

As mentioned above for a straight pipe, the flow and extra-stress profiles given in Eq. (4.5) satisfy the lubrication equations Eqs. (4.1)-(4.4a-d). However, the entrance extra-stress profiles given in Eq. (4.5) are not compatible with the exact solution at the wall given by Eq. (3.20). Indeed, from Eq. (4.5) we find that $\lim_{z \to 0^-} \dot{\gamma}(z) = -8$ and

$$\lim_{z \to 0^-} \tau_{zz}(1,z) = 128De, \quad \lim_{z \to 0^-} \tau_{yy}(1,z) = \lim_{z \to 0^-} \tau_{\theta\theta}(1,z) = 0, \quad \lim_{z \to 0^-} \tau_{yz}(1,z) = -8 \quad (4.8)$$

Since $\lim_{z \to 0^+} \dot{\gamma}(z) = -8$ and $\lim_{z \to 0^+} H'(z) \neq 0$, from Eqs. (3.19) we also find:

$$\lim_{z \to 0^+} \tau_{zz}(1,z) = 128De + \underline{16\varepsilon^2 H'(0)(1+8De H'(0))}$$
$$\lim_{z \to 0^+} \tau_{yy}(1,z) = -16H'(0) + 128De H'(0)^2 + \underline{128\,\varepsilon^2 De H'(0)^4}$$
$$\lim_{z \to 0^+} \tau_{\theta\theta}(1,z) = 0 \quad (4.9)$$
$$\lim_{z \to 0^+} \tau_{yz}(1,z) = -8 + 128De H'(0) + \underline{8\varepsilon^2 H'(0)^2(1+16DeH'(0))}$$

where in Eq. (4.9) the underlying $O(\varepsilon^2)$ terms are disregarded at the classic lubrication limit, i.e., for $\varepsilon^2 \to 0$, and have been included here for completeness only. Therefore, Eqs. (4.8) and (4.9) show that the viscoelastic extra-stress $\tau_{yz}$ and $\tau_{yy}$ at the inlet of the varying region of the pipe are discontinuous, and therefore a substantial rearrangement of the stresses is expected to take place when the fluid enters the varying region.

### 4.4. Pressure-drop

The general formula for the average pressure-drop required to drive the flow, at the classic lubrication limit, is found from Eq. (3.6) by taking into account that the leading-order pressure field is independent of the transverse direction *y*:

$$\Delta\Pi = -\int_0^1 P'(z)H^2(z)dz + \int_0^1 (H^2(z)-1)P'(z)dz = -\int_0^1 P'(z)dz \quad (4.10)$$

Although Eq. (4.10) may seem trivial, it is a consequence of the classic lubrication approximation according to which the pressure gradient is independent of the radial coordinate. For $Re_m=0$, we proceed by multiplying Eq. (4.2) with $y$, integrating with respect to $y$ from $y=0$ to $y=H(z)$ and applying the conditions at the axis of symmetry. Simplifying the result, gives:

$$P'(z) = \frac{2}{H}\left((1-\eta)\frac{\partial U}{\partial y} + \eta(\tau_{yz} - H'\tau_{zz})\right)_{y=H} + \frac{2\eta}{H^2}\frac{d}{dz}\left(\int_0^H \tau_{zz}\, y\, dy\right) \qquad (4.11)$$

Eq. (4.11) can be simplified further by considering Eq. (3.22) at the limit of a vanishing small aspect ratio, i.e., by omitting the $O(\varepsilon^2)$ term. In this case, $(\tau_{yz} - H'\tau_{zz})_{y=H} = \dot{\gamma}$, where $\dot{\gamma} = \partial U / \partial y|_{y=H}$ and thus, Eqs. (4.10) and (4.11) give:

$$\Delta\Pi = \underbrace{-2\int_0^1 \frac{\dot{\gamma}(z)}{H(z)}dz}_{\dot{\gamma}_w} + \underbrace{2\eta\left(I(0) - \Lambda^2 I(1) - \int_0^1 \frac{2H'(z)}{H^3(z)}I(z)dz\right)}_{\tau_w} \qquad (4.12)$$

In Eq. (4.12) $I(z) := \int_0^{H(z)} \tau_{zz}(y,z)\, y\, dy$, and due to the definitions shown in Eq. (4.12), $\Delta\Pi = \dot{\gamma}_w + \tau_w$, namely the average pressure drop along the pipe is a consequence of the tangential viscous stresses along the wall, $\dot{\gamma}_w$, plus an additional term that comes from the axial viscoelastic extra-stress along the main flow direction.

For the hyperbolic pipe described by Eq. (2.35), $2H'(z)/H^3(z) = 1 - \Lambda^2$ and Eq. (4.12) reduces to:

$$\Delta\Pi = -2\int_0^1 \frac{\dot{\gamma}(z)}{H(z)}dz + 2\eta\left(I(0) - \Lambda^2 I(1) - (1-\Lambda^2)\int_0^1 I(z)dz\right) \qquad (4.13)$$

For a Newtonian fluid, $\tau_{zz} = 0$ (see Eq. (4.4a)) and Eq. (4.12), or Eq. (4.13), gives:

$$\Delta\Pi_N = 16\int_0^1 \frac{dz}{H^4(z)} \qquad (4.14a)$$

For the hyperbolic pipe Eq. (4.14a) reduces to:

$$\Delta\Pi_N = 16(1 + \Lambda^2 + \Lambda^4)/3 \qquad (4.14b)$$

Finally, for a straight pipe, i.e., for $H(z)=1$, Eqs. (4.14a,b) give $\Delta\Pi_N = 16$.

### 4.5. First normal stress difference and Trouton ratio

Assuming that the polymer molecules are unstretched at the entrance of the varying region of the pipe, the general formula for the first normal stress difference at the lubrication limit is given directly from Eq. (3.18b) because as mentioned in § 3, $\varepsilon^2$ does not appear directly in Eq. (3.18b). Thus, $N_1$ can be found provided that the velocity profile at the axis of symmetry, $u = u(z)$, is known. This of course requires the solution of the full governing equations albeit approximation(s) of $u$ can be utilized too. For instance, for a Newtonian fluid ($De = 0$) and creeping flow ($\text{Re}_m = 0$), and using Eq. (4.6) to find the velocity at the axis of symmetry, $u_N(z) \equiv U_N(y = 0, z) = 4/H^2(z)$ and $u'_N(z) = -8H'(z)/H^3(z)$, reduces Eq. (3.18b) as follows

$$N_{1,N} = 3u'_N(z) = -24\, H'(z)/H^3(z) \tag{4.15}$$

The approximation of the velocity profile with the corresponding Newtonian one is a common approach in literature for hyperbolic geometries and many fluids with different rheological behavior such as viscoelastic Boger-type fluids or inelastic shear thinning-fluids (Ober *et al.* 2013; Perez-Salas et al. 2019; James & Tripathi 2023). Although the first normal stress difference shown in Eq. (4.14) appears to depend on the axial coordinate, due to the dependence of $H$ on $z$, for the hyperbolic pipe described by Eq. (2.35) the ratio $H'/H^3$ is constant and equals with $-(\Lambda^2 - 1)/2$ yielding a simple formula for $N_{1,N}$ which depends on the geometric ratio $\Lambda \equiv h_0^*/h_f^*$ only, i.e. $N_{1,N} = 12(\Lambda^2 - 1)$. Therefore, for $\Lambda \neq 1$, the Trouton ratio, defined as $Tr \equiv \dfrac{N_1}{u'(z)} = \dfrac{N_1}{4(\Lambda^2 - 1)}$, gives the well-established result for a Newtonian fluid under homogeneous uniaxial extension:

$$Tr_N = 3, \quad \Lambda \neq 1 \tag{4.16}$$

For the viscoelastic case, one can use the Newtonian velocity profile in Eq. (3.18b), to find an approximation of the Trouton ratio; this merely corresponds to the uncoupled case ($\eta = 0$) at the lubrication limit ($\varepsilon^2 \to 0$). Assuming the hyperbolic pipe, i.e., considering Eq. (2.35) for the shape function and substituting in Eq. (3.18b) yields $Tr = Tr(z, \eta, \Lambda, De_m)$:

$$Tr = 3(1-\eta) + \frac{\eta}{De_m}\left( \underbrace{\frac{1 - 2De_m\left(1 + (\Lambda^2 - 1)z\right)^{2 - \frac{1}{De_m}}}{1 - 2De_m}}_{C_{zz}(0,z)} - \underbrace{\frac{1 + De_m\left(1 + (\Lambda^2 - 1)z\right)^{-1 - \frac{1}{De_m}}}{1 + De_m}}_{C_{yy}(0,z)} \right) \tag{4.17}$$

where hereafter $\Lambda \neq 1$ and the modified Deborah number $De_m \equiv 4De(\Lambda^2 - 1)$ is used for brevity; for a contracting pipe, $\Lambda$ is larger than one and thus $De_m$ is positive. We do emphasize that the modified Deborah number plays an important role in the solution of the major quantities of interest. Note that the first term in parenthesis on the right-hand side of Eq. (4.17) is merely $C_{zz}(0,z)$, while the second is $C_{yy}(0,z) = C_{\theta\theta}(0,z)$; for completeness, we also report that $\varphi(z) = \left(1 + (\Lambda^2 - 1)z\right)^{-1/De_m}$. In contrast to the Trouton ratio for a Newtonian fluid, Eq. (4.16), the formula for a viscoelastic fluid, Eq. (4.17), depends on the axial coordinate, $z$. Evaluating Eq. (4.17) at the exit of the pipe, we find $Tr_{ex} \equiv Tr(1, \eta, \Lambda, De_m)$:

$$Tr_{ex} = 3(1-\eta) + \frac{\eta}{De_m}\left(\frac{1 - 2De_m \Lambda^{2(2-1/De_m)}}{1 - 2De_m} - \frac{1 + De_m \Lambda^{-2(1+1/De_m)}}{1 + De_m}\right) \tag{4.18}$$

For instance, in Figure 2a, Eq. (4.17) is shown as function of z for $De_m$=0.3 (black solid line), 0.5 (dashed red line) and 0.7 (dotted blue line) for Λ=3 and η=4/10. A monotonic increase of the Trouton ratio with the distance from the inlet is depicted, and the same holds with the increase of Λ. Also, Figure 2b shows the Trouton ratio at z=1, i.e., $Tr_{ex}$, as a function of $De_m$, for η=4/10 and Λ=2 (solid black line), 2.5 (dashed red line) and 3 (dotted blue line). A smooth enhancement of the Trouton ration is observed, although at high enough modified number a slight decrease appears.

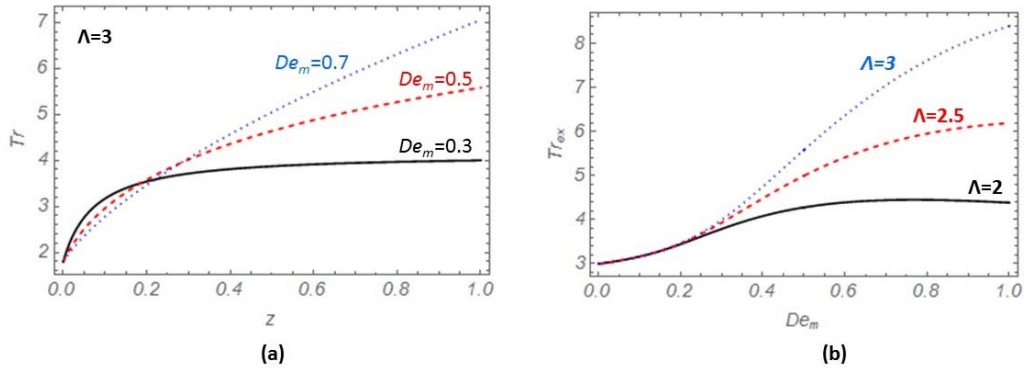

**Figure 2**: The Trouton ratio, Eq. (4.17), for a fluid with η=4/10 based on the Newtonian velocity profile at the classic lubrication limit (Eq. (4.6)):
(a) versus z; (b) evaluated at z=1 versus $De_m$

We also comment on the peculiar behavior of $C_{zz}(0,z)$ as $De_m \to 1/2$. In this case, $C_{zz}(0,z)$ is defined only as a limit which is given by:

$$\lim_{De_m \to 1/2} C_{zz}(0,z) = 1 + 2\ln\left(1 + (\Lambda^2 - 1)z\right) \tag{4.19}$$

We note however that $C_{zz}(0,z)$ is continuous and differentiable for any $De_m > 0$ (including $De_m = 1/2$). From Eqs. (4.17)-(4.19) we find the corresponding limits of $Tr$ and $Tr_{ex}$, respectively:

$$\lim_{De_m \to 1/2} Tr = 3 + \eta\left(4\ln\left(1 + (\Lambda^2 - 1)z\right) - \frac{2}{3\left(1 + (\Lambda^2 - 1)z\right)^3} - \frac{7}{3}\right) \tag{4.20}$$

and

$$\lim_{De_m \to 1/2} Tr_{ex} = 3 + \eta\left(8\ln(\Lambda) - \frac{2}{3\Lambda^6} - \frac{7}{3}\right) \tag{4.21}$$

Additional formulas for the Trouton ratio follow in § 5, while further comments and discussion are given in § 7.

## 5. High-order asymptotic solution and results

Instead of solving the Eqs (4.1)-(4.4) for the primary field variables $\{U, V, p, \tau_{zz}, \tau_{yz}, \tau_{yy}, \tau_{\theta\theta}\}$, we introduce the stream function, $\Psi$, which is defined with the aid of the two velocity components:

$$U = \frac{1}{y}\frac{\partial \Psi}{\partial y}, \quad V = -\frac{1}{y}\frac{\partial \Psi}{\partial z} \tag{5.1}$$

Thus, the continuity equation, Eq. (4.1), is satisfied automatically and the pressure gradient can be eliminated by differentiating Eq. (4.2) with respect to $y$. Thus, Eqs (4.1)-4.3) are replaced with the following equation:

$$\text{Re}_m \frac{\partial}{\partial y}\left(\frac{D}{Dt}\left(\frac{1}{y}\frac{\partial \Psi}{\partial y}\right)\right) = \frac{(1-\eta)}{y}\left(\frac{\partial^4 \Psi}{\partial y^4} - 2\frac{\partial^3 \Psi}{\partial y^3} + \frac{3}{y^2}\frac{\partial^2 \Psi}{\partial y^2} - \frac{3}{y^3}\frac{\partial \Psi}{\partial y}\right) + \\ +\eta\left(\frac{\partial^2 \tau_{zz}}{\partial z \partial y} + \frac{\partial^2 \tau_{yz}}{\partial y^2} + \frac{1}{y}\frac{\partial \tau_{yz}}{\partial y} - \frac{\tau_{yz}}{y^2}\right) \tag{5.2}$$

where

$$\frac{\partial}{\partial y}\left(\frac{D}{Dt}\left(\frac{1}{y}\frac{\partial \Psi}{\partial y}\right)\right) = \frac{1}{y^2}\left(\left(\frac{3}{y^2}\frac{\partial \Psi}{\partial z} - \frac{1}{y}\frac{\partial \Psi}{\partial y}\frac{\partial \Psi}{\partial z} + \frac{\partial^3 \Psi}{\partial y \partial z^2}\right)\frac{\partial \Psi}{\partial y} + \left(\frac{3}{y}\frac{\partial^2 \Psi}{\partial y^2} - \frac{\partial^3 \Psi}{\partial y^3}\right)\frac{\partial \Psi}{\partial z}\right) \tag{5.3}$$

Since both $U$ and $V$ (and their gradients) are expressed in terms of derivatives of the stream function only, a reference value for $\Psi$ can be chosen arbitrary. Substituting $U$ in the total mass balance, Eq. (2.36d), gives $\Psi(H(z), z) - \Psi(0, z) = 1$ which indicates that one of $\Psi(0, z)$ and $\Psi(H(z),z)$ can be set conveniently. Also, considering the boundary conditions at the wall and the restrictions needed so that the governing equations are well-defined at the axis of symmetry of the pipe, the final conditions for the stream function are summarized as follows:

$$\Psi(0,z) = \frac{\partial \Psi}{\partial y}(0,z) = \frac{\partial^3 \Psi}{\partial y^3}(0,z) = 0, \quad \Psi(H(z),z) = 1, \frac{\partial \Psi}{\partial z}(H(z),z) = \frac{\partial \Psi}{\partial y}(H(z),z) = 0 \quad (5.4)$$

For a Newtonian fluid (*De*=0) and creeping flow conditions ($\text{Re}_m = 0$) the analytical solution for the stream function in the varying region of the pipe, $0 \le z \le 1$, is*:*

$$\Psi_N = \frac{y^2}{H^2}\left(2 - \frac{y^2}{H^2}\right) \quad (5.5)$$

It can be trivially confirmed that Eq. (5.5) satisfies all conditions given in (5.4). At the entrance region, $z \le 0$, the shape function is constant, $H(z)=1$, and Eq. (5.5) reduces to $\Psi_{en} = y^2(2-y^2)$ where we have used the subscript "*en*" in accordance with Eq. (4.5).

For the solution of Eqs. (5.2) and (4.4a-d) we assume a regular perturbation scheme in terms of the Deborah number:

$$f \approx f_N + \sum_{k=1}^{\infty} f_k De^k, \quad 0 < De \ll 1 \quad (5.6)$$

where $f = \Psi, \tau_{zz}, \tau_{yy}, \tau_{yz}$ and $\tau_{\theta\theta}$. The series are substituted into the governing equations and accompanied boundary conditions yielding a sequence of equations which are solved analytically up to $O(De^8)$ with the aid of the "MATHEMATICA" software (Wolfram 2023). This quite demanding task requires advanced techniques and substantial computer resources to be achieved; it is also more difficult than that for the planar case solved and presented recently by Housiadas & Beris (2023). The $O(De^0) = O(1)$ equations correspond to the equations for a simple Newtonian fluid, while the effect of viscoelasticity is built with the addition of the $O(De^k)$ terms, *k*=1, 2,…, 8. Worth mentioning is also the fact that with this method of solution, inlet (at *z*=0) or outlet (at *z*=1) conditions for the stream function and the viscoelastic extra-stresses are not needed and cannot be imposed. Inlet conditions, however, are required when the equations are solved numerically; more details are given in § 6. The $O(De)$ solution is*:*

$$\Psi_1 = 0, \quad \tau_{zz,1} = \frac{128}{H^6}\bar{y}^2, \quad \tau_{yz,1} = 128\frac{H'}{H^6}\bar{y}(2\bar{y}^2-1),$$

$$\tau_{yy,1} = \frac{32}{H^5}\left(\frac{5H'^2}{H}-H''\right) + \frac{64}{H^5}\left(H''-\frac{9H'^2}{H}\right)\bar{y}^2 + \frac{32}{H^5}\left(\frac{17H'^2}{H}-H''\right)\bar{y}^4, \quad (5.7)$$

$$\tau_{\theta\theta,1} = \frac{32}{H^5}\left(\frac{5H'^2}{H}-H''\right) + \frac{64}{H^5}\left(H''-\frac{5H'^2}{H}\right)\bar{y}^2 + \frac{32}{H^5}\left(\frac{5H'^2}{H}-H''\right)\bar{y}^4$$

where we have used the normalized radial coordinate (or similarity variable) $\bar{y} = y/H(z)$ for brevity. Since $\Psi_1 = 0$, the $O(De)$ velocity field is zero as previously predicted by Boyko & Stone (2022) and Housiadas & Beris (2023) for the planar case, and according to the theorem of Tanner & Pipkin (1969). The $O(De^2)$ solution is:

$$\Psi_2 = \frac{\eta\,\bar{y}^2}{H^6}\left(\hat{\Psi}_2^{(2)}(z) + \bar{y}^2\,\hat{\Psi}_2^{(4)}(z) + \bar{y}^4\,\hat{\Psi}_2^{(6)}(z) + \bar{y}^6\,\hat{\Psi}_2^{(8)}(z)\right) \quad (5.8a)$$

$$\tau_{zz,2} = 3072\frac{H'}{H^9}\bar{y}^2(1-\bar{y}^2) \quad (5.8b)$$

$$\tau_{yz,2} = \frac{256}{H^9}\bar{y}\left(\hat{\tau}_{yz,2}^{(1)}(z) + \bar{y}^3\,\hat{\tau}_{yz,2}^{(3)}(z) + \bar{y}^5\,\hat{\tau}_{yz,2}^{(5)}(z)\right) \quad (5.8c)$$

$$\tau_{yy,2} = \frac{128}{H^9}\left(\hat{\tau}_{yy,2}^{(0)}(z) + \bar{y}^2\,\hat{\tau}_{yy,2}^{(2)}(z) + \bar{y}^4\,\hat{\tau}_{yy,2}^{(4)}(\bar{z}) + \bar{y}^6\,\hat{\tau}_{yy,2}^{(6)}(z)\right) \quad (5.8d)$$

$$\tau_{\theta\theta,2} = \frac{128}{H^9}\left(\hat{\tau}_{\theta\theta,2}^{(0)}(z) + \bar{y}^2\,\hat{\tau}_{\theta\theta,2}^{(2)}(z) + \bar{y}^4\,\hat{\tau}_{\theta\theta,2}^{(4)}(z) + \bar{y}^6\,\hat{\tau}_{\theta\theta,2}^{(6)}(z)\right) \quad (5.8e)$$

where $\hat{\Psi}_2^{(j)}$, $\hat{\tau}_{yz,2}^{(j)}$, $\hat{\tau}_{zz,2}^{(j)}$ and $\hat{\tau}_{\theta\theta,2}^{(j)}$ are functions of the shape function and its derivative(s) with respect to the axial coordinate; their precise form is provided in the Appendix. The higher-order solutions are too long to be printed here.

### 5.1. Average pressure-drop

Based on the series expansion in terms of *De*, the solution for the pressure gradient is:

$$P'(\bar{z}) \approx P'_N(\bar{z}) + \eta\sum_{k=1}^{8}\hat{P}'_k(\bar{z})De^k \quad (5.9)$$

where $\eta\hat{P}'_k(\bar{z}) \equiv P'_k(z)$, k=1,2,…,8. The individual components are provided below up to $O(De^4)$, while the higher-order components are too long to be printed here:

$$P'_N = -\frac{16}{H^4} \quad (5.10a)$$

$$\hat{P}'_1 = -256\frac{H'}{H^7} = \frac{128}{3}\left(\frac{1}{H^6}\right)' \quad (5.10b)$$

$$\hat{P}_2' = 768\left(\frac{H''}{H^9} - \frac{19}{3}\frac{H'^2}{H^{10}}\right) \tag{5.10c}$$

$$\hat{P}_3' = 3072\left(-\frac{152}{3}\frac{H'^3}{H^{13}} + \frac{296}{15}\frac{H'H''}{H^{12}} - \frac{16}{15}\frac{16H'''}{15H^{11}} + \frac{\eta}{5}\left(52\frac{H'^3}{H^{13}} - \frac{59}{3}\frac{H'H''}{H^{12}} + \frac{H'''}{H^{11}}\right)\right) \tag{5.10d}$$

$$\hat{P}_4' = -\frac{840H'^4}{H^{16}} + \frac{1586H'^2H''}{3H^{15}} - \frac{85H''^2}{3H^{14}} - \frac{46H'H'''}{H^{14}} + \frac{5H^{(4)}}{3H^{13}} + \eta\left(\frac{9221H'^4}{20H^{16}} - \frac{2729H'^2H''}{10H^{15}} + \frac{109H''^2}{8H^{14}} + \frac{133H'H'''}{6H^{14}} - \frac{89H^{(4)}}{120H^{13}}\right) + \eta^2\left(-\frac{136H'^4}{H^{16}} + \frac{1178H'^2H''}{15H^{15}} - \frac{11H''^2}{3H^{14}} - \frac{94H'H'''}{15H^{14}} + \frac{H^{(4)}}{5H^{13}}\right) \tag{5.10e}$$

where the explicit dependence on the axial coordinate is omitted for simplicity. For a hyperbolic pipe, i.e., when the shape function is given by Eq. (2.35), Eq. (4.2) gives up to eight-order:

$$\frac{\Delta\Pi}{\Delta\Pi_N} \approx 1 - 2\eta\, De_m + \frac{5\eta}{2}De_m^2 + \eta\left(-\frac{14}{5} + \frac{3\eta}{5}\right)De_m^3 + \eta\left(3 - \frac{49}{20}\eta + \frac{4}{5}\eta^2\right)De_m^4 + \sum_{k=5}^{8}\eta\,\delta_k(\eta)De_m^k \tag{5.11}$$

where $\Delta\Pi_N$ is given in Eq. (4.14b), and $\delta_k = \delta_k(\eta)$, $k=5,6,7,8$, are given by:

$$\delta_5(\eta) = -\frac{22}{7} + \frac{653}{105}\eta - \frac{506}{105}\eta^2 + \frac{8}{7}\eta^3 \tag{5.12a}$$

$$\delta_6(\eta) = \frac{13}{4} - \frac{4399}{350}\eta + \frac{9049}{525}\eta^2 - \frac{4849}{525}\eta^3 + \frac{12}{7}\eta^4 \tag{5.12b}$$

$$\delta_7(\eta) = -\frac{10}{3} + \frac{4649}{210}\eta - \frac{16609}{350}\eta^2 + \frac{270569}{6300}\eta^3 - \frac{5512}{315}\eta^4 + \frac{8}{3}\eta^5 \tag{5.12c}$$

$$\delta_8(\eta) = \frac{17}{5} - \frac{6397}{180}\eta + \frac{2095259}{18900}\eta^2 - \frac{56720593}{378000}\eta^3 + \frac{791519}{7875}\eta^4 - \frac{259843}{7875}\eta^5 + \frac{64}{15}\eta^6 \tag{5.12d}$$

It is interesting, and somehow unexpected, that the final series for the reduced average pressure drop can be recast in terms of $De_m \equiv 4(\Lambda^2 - 1)De$. Furthermore, an amazing feature of formula (5.11) is that is independent of the contraction ratio Λ. A similar formula which does not depend on Λ can also be derived for the planar case, studied theoretically previously by Boyko & Stone (2022) and Housiadas & Beris (2023), albeit these authors did not notice and did not report the corresponding formula.

To get more insight about the results, we consider *η*=4/10 and *η*=1 and we calculate the coefficients of the $O(De_m^k)$ terms, $k=0,1,2,\ldots,8$, in Eq. (5.11); the results are reported in Table 1. We notice that the numerical coefficients for *η*=0.4 are smaller than those for *η*=1, as well as that the magnitude of the coefficients decreases very slowly. Thus, and to increase the accuracy of our eight-order formula, Eq. (5.11), we proceed by applying a technique that rearranges non-linearly the terms of a truncated series. Specifically, we apply the diagonal [M/M] Padé approximants (Padé 1892) on Eq. (5.11), where M=1, 2, 3 and 4. and we derive new, transformed, analytical solutions. We remind the reader that the diagonal [M/M] Padé approximant of a function $f=f(\delta)$ agrees with the corresponding Taylor series of $f$ about the point $\delta=0$ up to $O(\delta^{2M})$. The successive approximants for M=1,2,3 and 4 can be used to check the convergence of the approximants, namely the convergence of the transformed solutions.

| $\dfrac{\Delta\Pi}{\Delta\Pi_N}$ | $O(1)$ | $O(De_m)$ | $O(De_m^2)$ | $O(De_m^3)$ | $O(De_m^4)$ | $O(De_m^5)$ | $O(De_m^6)$ | $O(De_m^7)$ | $O(De_m^8)$ |
|---|---|---|---|---|---|---|---|---|---|
| $\eta=0.4$ | 1 | -0.8 | 1 | -1.024 | 0.859 | -0.541 | 0.173 | 0.103 | -0.171 |
| $\eta=1$ | 1 | -2 | 2.5 | -2.2 | 1.35 | -0.6 | 0.396 | -0.534 | 0.448 |

**Table 1:** The coefficients of the $O(De_m^k)$ terms, $k=0,1,2,...,8$, in Eq. (5.11)

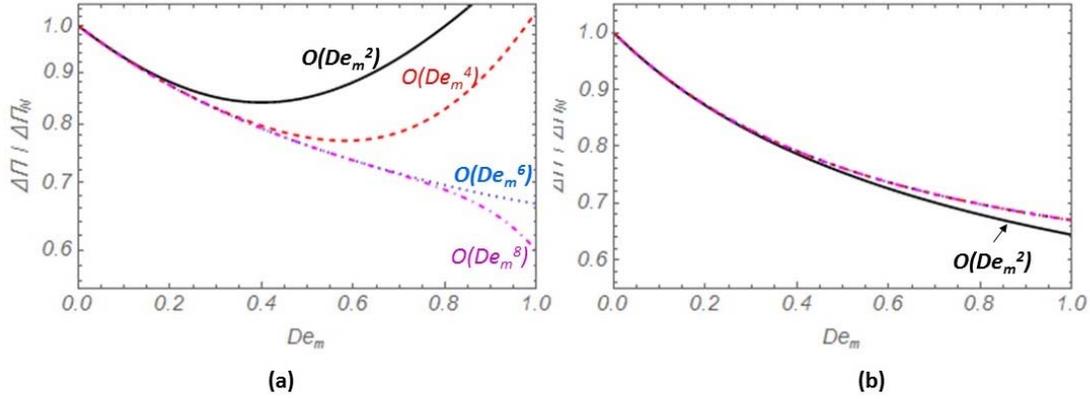

**Figure 3:** Reduced pressure drop vs $De_m$ for the Oldroyd-B model with $\eta$=0.4
(a) Perturbation solutions:
solid (black) line: 2nd order solution; dashed (red) line: 4th order solution;
dotted (blue) line: 6th order solution; dot-dashed (magenta) line: 8th order solution
(b) Accelerated solutions:
solid (black) line: up to 2nd order; dashed (red) line: up to 4th order;
dotted (blue) line: up to 6th order; dot-dashed (magenta) line: up to 8th order

In Figure 3a, we present the reduced pressure drop, $\Delta\Pi/\Delta\Pi_N$, for $\eta=4/10$ up to second-, fourth-, sixth and eight-order perturbation solutions as functions of $De_m$. Convergence of the results is clearly observed gradually as more terms are included in the series, and we safely claim that the radius of convergence of the series is at least 0.85. The results show a decrease of $\Delta\Pi/\Delta\Pi_N$, as previously predicted for a hyperbolic symmetric channel (Boyko & Stone 2022; Housiadas & Beris 2023). Furthermore, the accelerated (transformed) solutions which are depicted in Figure 3b, clearly show the convergence of the perturbation results. The convergence is achieved when at least five terms in the series are taken into account for the construction of the Padé [M/M] diagonal approximant (i.e., for M=2). Indeed, the curves with M=2, 3 and 4, which are constructed using the first five, seven and nine terms in the series (5.11), respectively, are practically indistinguishable.

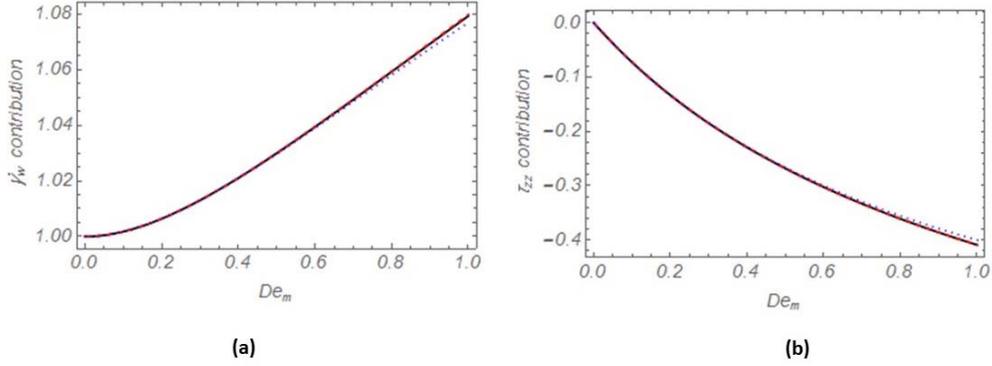

**Figure 4:** Decomposition of the average pressure drop, ΔΠ, based on Eq. (4.12), or Eq. (4.13) in (a) wall shear stress contribution and (b) viscoelastic normal extra-stress contribution
The contributions are normalized by $\Delta\Pi_N$ (Eq. (4.14b)) and the viscosity ratio is η=4/10.
Solid (black), dashed (red) and dotted (blue) lines are acceleration formulas up to $O(De^8)$, $O(De^6)$ and $O(De^4)$, respectively

In Figure 4, we present the individual contributions to the average pressure drop, i.e., $\dot{\gamma}_w$ and $\tau_w$ (see Eq. (4.12)) normalized by the Newtonian value $\Delta\Pi_N$, as function of the modified Deborah number for a polymer viscosity ratio η=4/10. First, we notice the clear convergence of both contributions to the average pressure drop. We also see that with increasing $De_m$, an increase of the magnitude of the positive viscous contribution at the wall ($\dot{\gamma}_w > 0$) is predicted, as well as the increase of the magnitude of the negative viscoelastic contribution ($\tau_w < 0$). Interestingly, however, the negative viscoelastic contribution overwhelms the positive viscous contribution leading to a decrease in the average pressure drop. Similar results were observed for the planar case too (Housiadas & Beris, 2023). For completeness, the pure viscous and viscoelastic contributions are given up to fourth order in $De_m$:

$$\frac{\dot{\gamma}_w}{\Delta\Pi_N} \approx 1 + \frac{\eta}{2}De_m^2 + \frac{\eta}{5}(3\eta - 4)De_m^3 + \eta\left(1 + \frac{\eta}{20}(16\eta - 37)\right)De_m^4 \quad (5.13)$$

and

$$\frac{\tau_w}{\Delta \Pi_N} \approx 2\eta De_m \left( -1 + De_m - De_m^2 + \left(1 - \frac{3\eta}{10}\right) De_m^3 \right) \tag{5.14}$$

### 5.2. First normal stress difference and Trouton ratio

As far as the exact solution for the Trouton ratio is concerned, given by Eq. (3.18b), we need to emphasize that the regular perturbation scheme in terms of *De* employed above cannot be applied directly on Eq. (3.18b). The reason is that the function $\varphi(z) = \exp\left(-De^{-1} \int_0^z u^{-1}(x) dx\right)$ which appears in (3.18a,b) goes to zero faster than any power of the Deborah number, namely $\varphi = o(De^n)$ for any $n = 0, 1, 2, 3, \ldots$ as $De \to 0^+$. Provided that the fluid velocity at the axis of symmetry, $u = u(z)$, is a continuous and non-zero positive function for any $z \in [0,1]$, $\varphi$ is smooth and zero over an infinitely long interval, i.e. for $0 < De < \infty$, and yet nonzero, but it cannot be described by a Taylor series because it is not holomorphic.

However, and even if we are aware that the exact solution cannot be approximated by a simple power series in terms of *De*, we pretend that we do not know this information, and we proceed by going one step back, i.e., we start from Eq. (3.9ba) and apply the lubrication expansion:

$$N_1 = 3(1-\eta) u'_{(0)} + \frac{\eta}{De} \left( \frac{c_{zz(0)}}{\varepsilon^2} + c_{zz(2)} - c_{yy(0)} \right) + O(\varepsilon^2) \text{ at } y = 0 \tag{5.15}$$

where we retain the lubrication subscript to avoid confusion, $u'_{(0)} \equiv \partial U_{(0)} / \partial z \big|_{y=0}$ and the equations that govern $c_{zz(0)}(0,z)$, $c_{zz(2)}(0,z)$ and $c_{yy(0)}(0,z)$ are found from Eqs. (3.10) and (3.11). For completeness, we give the leading-order fluid velocity at the axis of symmetry, $u_{(0)}$, which, interestingly, can be recast as follows:

$$u_{(0)} = u_{(0)N}(z) q_{[n]}(\eta, De_m) \tag{5.16a}$$

$$q_{[n]} = 1 + \hat{q}_2(\eta) De_m^2 + \ldots + \hat{q}_n(\eta) De_m^n + O(De_m^{n+1}) \tag{5.16b}$$

where $u_{(0)N} = 4/H^2(z) = 4(1 + (\Lambda^2 - 1)z)$, and therefore $u'_{(0)} = 4(\Lambda^2 - 1) q_{[n]}(\eta, De_m)$. The first three functions $\hat{q}_k$, $k = 2, 3, 4$ are:

$$\hat{q}_2 = -\frac{\eta}{6}, \quad \hat{q}_3 = \frac{\eta}{30}(11 - 7\eta), \quad \hat{q}_4 = -\frac{\eta}{300}\left(170 + \eta(104\eta - 283)\right) \tag{5.17}$$

First, the initial value problem that determines $c_{zz(0)}(0,z)$ is:

$$c_{zz(0)} + De\left(u_{(0)}\, c'_{zz(0)} - 2c_{zz(0)}u'_{(0)}\right) = 0, \quad 0 < z \leq 1$$
$$c_{zz(0)}(0, z=0) = 0 \tag{5.18}$$

where an unstretched inlet condition for the polymer molecules is imposed. The perturbation solution of Eq. (5.18) in terms of *De* (or *De$_m$*), up to any order, is:

$$c_{zz(0)}(0,z) = 0, \quad 0 \leq z \leq 1 \tag{5.19}$$

Using Eq. (5.19), we find that the initial value problem that determines $c_{zz(2)}(0,z)$ is:

$$c_{zz(2)} + De\left(u_{(0)}\, c'_{zz(2)} - 2c_{zz(2)}u'_{(0)}\right) = 1, \quad 0 < z \leq 1$$
$$c_{zz(2)}(0, z=0) = 1 \tag{5.20}$$

Unexpectedly, we see that the solution for $c_{zz(2)}(0,z)$ can be found using only $u_{(0)}$. The solution of Eq. (5.20) up to fourth order in $De_m$ is:

$$c_{zz(2)} = 1 + 2De_m + 4De_m^2 + \left(8 - \frac{\eta}{3}\right)De_m^3 + \left(16 - \frac{1}{15}\eta(9+7\eta)\right)De_m^4 + O(De_m^5) \tag{5.21}$$

Similarly, the initial value problem that determines $c_{yy(0)}(0,z)$ is:

$$c_{yy(0)} + De\left(u_{(0)}\, c'_{yy(0)} + c_{yy(0)}u'_{(0)}\right) = 1, \quad 0 < z \leq 1$$
$$c_{yy(0)}(0, z=0) = 1 \tag{5.22}$$

The solution for $c_{yy(0)}(0,z)$ up to fourth order in $De_m$ is:

$$c_{yy(0)} = 1 + De_m^2 + \frac{1}{6}(\eta - 6)De_m^3 + \frac{1}{30}(30 + 7(\eta-3)\eta)De_m^4 + O(De_m^5) \tag{5.23}$$

We reiterate that $c_{zz(0)}$, $c_{zz(2)}$, $c_{yy(0)}$ and $u'_{(0)N}$ are independent of *z*. Based on Eq. (5.15), the leading order first normal stress difference is:

$$N_{1(0)} = 3(1-\eta)u'_{(0)} + \frac{\eta}{De}(c_{zz(2)} - c_{yy(0)}) \tag{5.24}$$

Plugging all known quantities ($c_{zz(2)}$, $c_{yy(0)}$ and $u'_{(0)}$) in Eq. (5.24) and simplifying the result we find the Trouton ratio at the lubrication limit, $Tr \equiv N_{1(0)}/u'_{(0)}$:

$$Tr \approx 3\left(1 + \eta De_m + \frac{17}{6}\eta De_m^2 + \frac{\eta}{30}(161-17\eta)De_m^3 + \frac{\eta}{300}(3130 + 53\eta - 244\eta^2)De_m^4 + O(De_m^5)\right) \tag{5.25}$$

As previously observed for the reduced pressure-drop, given by Eq. (5.11), the perturbation solution for the Trouton ratio is expressed in terms of $De_m$ and $\eta$ only. For the UCM model, i.e., for *η*=1, we find:

$$Tr \approx 3\left(1 + De_m + \frac{17}{6}De_m^2 + \frac{24}{5}De_m^3 + \frac{2939}{300}De_m^4 + \frac{5647}{300}De_m^5 + \frac{1621259}{44100}De_m^6 + \right. \\ \left. \frac{6297799}{88200}De_m^7 + \frac{1103656411}{7938000}De_m^8\right) \quad (5.26)$$

In contrast to Eq. (5.11), all terms in Eqs. (5.25) and (5.26) are positive which is indicative of the existence of an irregularity at a critical value of $De_m$. Also, and unlike the coefficients of Eq. (5.11), we see that the magnitude of the coefficients of $De_m$ in Eqs. (5.25) and (5.26) increases. Applying the lowest order diagonal [M/M] Padé approximant in the truncated series (5.25) (the case with M=1), shows that the approximant becomes singular at the critical value $De_{m,c} = 6/17$. Increasing the parameter M of the Padé approximant does not eliminate the singular point but pushes it towards $1/2$; this implies that we cannot expect accurate predictions as $De_m = 1/2$ is approached.

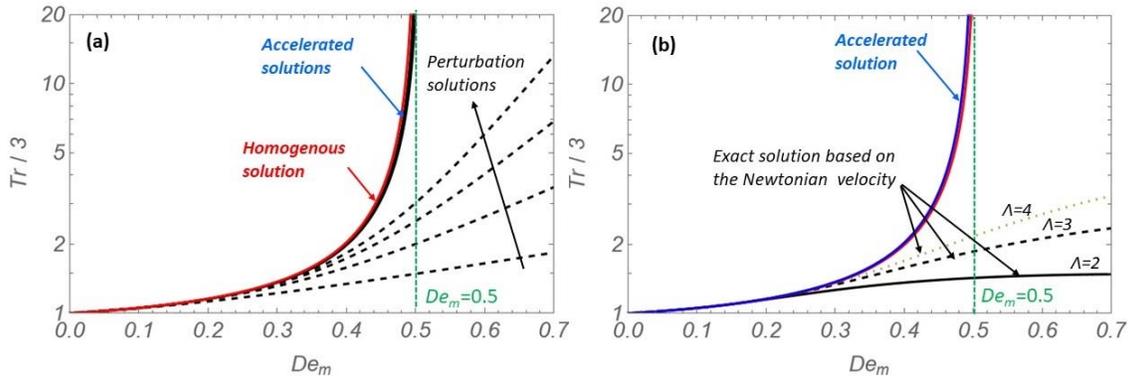

**Figure 5**: One third of the Trouton ratio as function of $De_m=4(\Lambda^2-1)De$ for $\eta=4/10$. In Fig. 5a, the perturbation solutions are calculated up to $O(De^2)$, $O(De^4)$, $O(De^6)$ and $O(De^8)$ (dashed lines respectively), while the accelerated solutions are constructed based on the perturbation solutions up to $O(De^4)$ and $O(De^8)$. The exact solutions shown in Fig. (5b) are based on the Newtonian velocity at the exit of the pipe.

In Figure 5, we present the Trouton ratio as predicted by Eq. (5.25) and increasing gradually the number of terms that are included in the series (shown with dashed lines), the corresponding Padé approximants with M=2, 3 and 4 (show with solid lines), along with the solution for the homogenous case (dotted red line):

$$Tr_h = 3\left(1 + \frac{\eta(1+2De_m)De_m}{(1-2De_m)(1+De_m)}\right) \quad (5.27)$$

Eq. (5.27) is derived assuming that there are not any effects from the boundaries and that the flow has reached at steady state, namely a pure steady uniaxial elongation is imposed (Bird, Armstrong & Hassanger 1987; Tanner 2000; Housiadas 2017). First, we observe that the Padé approximants are indistinguishable, as well as that they diverge near $De_m \approx 1/2$. Second, we see that the homogenous solution is practically the same as the transformed/accelerated solutions (i.e. with the Padé approximants for M=2,3 and 4). Third, we see that the perturbation solutions converge slowly and only for $De_m < ½$. This information in conjunction with the convergence of the successive Padé approximants implies that the radius of convergence of the perturbation series for the Trouton ratio, Eq. (5.25), is less than one half. Moreover, in Figure 5b we compare the accelerated solution up to $O(De_m^8)$ with the exact solution based on the Newtonian velocity $u_{(0)N} \equiv U_N(0,z)$, i.e. Eq. (4.18), for different values of the contraction ratio Λ; recall that Eq. (5.25) does not depend on Λ. It is seen that for $De_m > 0.3$ approximately, the accelerated solution cannot follow closely the exact solution as it should be expected due to its divergence near $De_m = 1/2$.

In Figure 6, we investigate the sensitivity of the Trouton ratio, $Tr \equiv N_1/u'(z)$ where $N_1$ is given by Eq. (3.18b), on the velocity profile at the axis of symmetry. Reiterating that Eq. (3.18b) is exact, we consider two cases, both of which have the same base velocity, i.e. the Newtonian velocity profile at the classic lubrication limit. In the first case, we consider higher order corrections due to viscoelasticity, namely in terms of the modified Deborah number, and in the second we consider higher order corrections due to variations in geometry, namely in terms of the square of the aspect ratio.

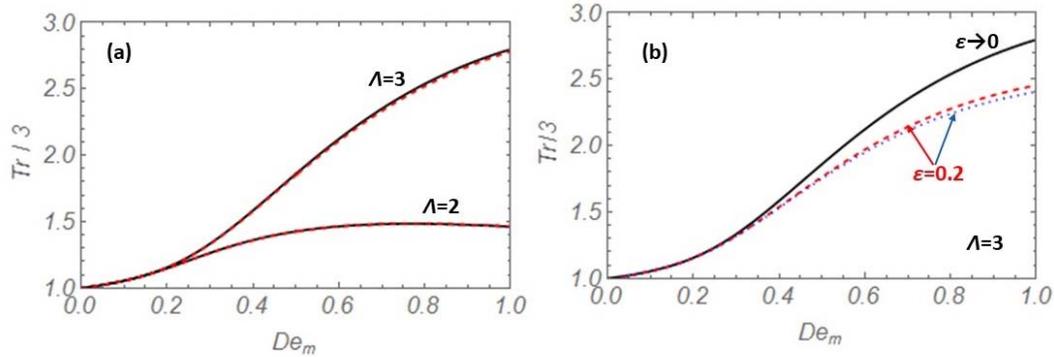

**Figure 6:** The Trouton ratio evaluated at z=1 as function of $De_m$ = 4(Λ²-1)De for η=4/10, using
(a) the Newtonian velocity, Eq. (4.6): solid (black) lines;
   the viscoelastic velocity profile up to $O(De^8)$ with acceleration: dashed (red) lines,
(b) the Newtonian velocity, Eq. (4.6): solid (black) line; the Newtonian velocity profile up to $O(ε^4)$ (Eq.(5.30));
   the Newtonian velocity profile up to $O(ε^4)$ with Padé [1/1] acceleration: dotted (blue) line

In the first case, the starting point of the analysis is Eq. (5.16a), $u_{(0)} \approx u_{(0)N}(z) q_{[2M]}(\eta, De_m)$, and its corresponding transformed formula based on the Padé [M/M] approximant, $u_{(0)} \approx u_{(0)N}(z) q_{[M/M]}(\eta, De_m)$ where $M = 1, 2, 3, 4$; recall that $q_{[M/M]}(\eta, De_m)$ agrees with $q_{[2M]}(\eta, De_m)$ up to $O(De_m^{2M})$. When these two approximate velocities are substituted into Eq. (3.18b), we find:

$$Tr = 3(1-\eta) + \frac{\eta}{a}\left( \frac{1 - 2a\left(1 + (\Lambda^2 - 1)z\right)^{2-\frac{1}{a}}}{1 - 2a} - \frac{1 + a\left(1 + (\Lambda^2 - 1)z\right)^{-1-\frac{1}{a}}}{1 + a} \right) \quad (5.28)$$

where

$$a \equiv De_m \, q_{[2M]}(\eta, De_m) \quad \text{or} \quad a \equiv De_m \, q_{[M/M]}(\eta, De_m) \quad (5.29)$$

namely, the first expression for $a$ is based on the truncated perturbation series up to $O(De_m^{2M})$, while the second expression is based on the corresponding Padé [M/M] approximant. Notice that Eq. (5.28) has the same form with Eq. (4.17) with the only difference that $a$ is in place of $De_m$. The Trouton ratio as function of $De_m$ is depicted in Figure 8a for $\eta$=4/10 and $\Lambda$=2 and 3. Clearly, the differences between the results are inconsequential.

In the second case, we use the Newtonian velocity profile at the axis of symmetry, $u_N$, up to fourth order in $\varepsilon$, as previously found by Housiadas & Tsagaris (2022), as well as the corresponding Padé [1/1] approximant with respect to $\varepsilon^2$:

$$u_N \approx u_{(0)} + \varepsilon^2 u_{(2)} + \varepsilon^2 u_{(4)}, \quad u_N \approx u_{(0)} + \frac{\varepsilon^2 u_{(2)}^2}{u_{(2)} - \varepsilon^2 u_{(4)}} \quad (5.30)$$

For the hyperbolic pipe, the individual contributions in Eq. (5.30) are:

$$u_{(0)} = 4(1 + (\Lambda^2 - 1)z)$$

$$u_{(2)} = \frac{(\Lambda^2 - 1)^2}{6\left(1 + (\Lambda^2 - 1)z\right)^2} \quad (5.31)$$

$$u_{(4)} = -\frac{(\Lambda^2 - 1)^2 \left(9 + 27(\Lambda^2 - 1)z + (\Lambda^2 - 1)^2(27z^2 - 2) + 9(\Lambda^2 - 1)^3 z^3\right)}{72\left(1 + (\Lambda^2 - 1)z\right)^5}$$

Using the velocity profiles shown in Eqs. (5.30)-(5.31), and $\eta$=4/10, $\Lambda$=3, we present the results for the Trouton ratio as function of the modified Deborah number in Figure 6b. The results are shown at the classic lubrication, $\varepsilon^2 \to 0$, and for the high aspect ratio $\varepsilon = 0.2$; note that for $\varepsilon \leq 0.1$ no observable differences are predicted. It is seen that the as the aspect ratio increases, the Trouton ratio drops, as well as that the differences between $u_N$ and the corresponding Padé approximant cause very small differences on the Trouton ration too.

Based on the sensitivity analysis performed here and the results presented in Figures 5 and 6, we conclude that the exact analytical solution for the Trouton ratio, which is extracted with the aid of Eq. (3.18b) and is based on the velocity at the axis of symmetry of the pipe, is insensitive to small changes from the Newtonian velocity field, and therefore the approximate analytical solution at the classic lubrication limit, i.e. Eq. (4.18), can safely be used for viscoelastic fluids in axisymmetric hyperbolic pipes with $\varepsilon \leq 0.1$. More comments and discussion on the theoretical features of the perturbation solution can be found in § 7.

## 6. Numerical solution of the lubrication equations

Eqs. (5.2) and (4.4a-d) consist of an initial (in the axial direction, $z$) and boundary (in the radial direction, $y$) value problem accompanied with the boundary conditions given in Eq. (5.4) and appropriate initial conditions. For the numerical solution of this system, first we introduce new coordinates $(\xi, \zeta)$ that map the varying boundary of the flow domain into a fixed one:

$$\xi = -1 + \frac{2y}{H(z)}, \quad \zeta = z \quad \Leftrightarrow \quad y = \frac{\xi+1}{2} H(\zeta), \quad z = \zeta \tag{6.1}$$

Thus, the shape function $H$ appears into the governing equations and the domain of definition of the lubrication equations becomes $\bar{\Omega} = \{(\xi,\zeta) \mid -1 < \xi < 1, \, 0 < \zeta \leq 1\}$. Based on Eq. (6.1) the first order partial differential operators are transformed as follows:

$$\frac{\partial}{\partial y} = \frac{2}{H(\zeta)} \frac{\partial}{\partial \xi}, \quad \frac{\partial}{\partial z} = \frac{\partial}{\partial \zeta} - \frac{(1+\xi)H'(\zeta)}{H(\zeta)} \frac{\partial}{\partial \xi} \tag{6.2}$$

where we have used $\partial H / \partial y = 0$ to show that $H = H(\zeta)$ only. Using Eq. (6.2) we find the higher-order derivatives with respect to $y$ and/or $z$ in terms of derivatives with respect to $\xi$ and $\zeta$. Substituting the transformed derivatives and the rescaled components of the conformation tensor into Eqs. (5.2) and (4.4a-d) gives the final equations for $\Psi, c_{zz}, c_{yy}$ and $c_{yz}$ defined on $\bar{\Omega}$; notice that we prefer to solve for the conformation tensor components instead of the components of the viscoelastic extra-stress tensor. Based on the stream function and the new coordinates, the material derivative is transformed as follows:

$$\frac{D}{Dt} \equiv \frac{4}{(1+\xi)H^2} \left( \frac{\partial \Psi}{\partial \xi} \frac{\partial}{\partial \zeta} - \frac{\partial \Psi}{\partial \zeta} \frac{\partial}{\partial \xi} \right) \tag{6.3}$$

Similarly, the velocity components are given as:

$$U = \frac{4}{(1+\xi)H^2}\frac{\partial \Psi}{\partial \xi}, \quad V = \frac{2}{H}\left(\frac{H'}{H}\frac{\partial \Psi}{\partial \xi} - \frac{1}{(1+\xi)}\frac{\partial \Psi}{\partial \zeta}\right) \qquad (6.4)$$

Finally, the velocity gradients are:

$$\frac{\partial U}{\partial z} = \frac{4}{(1+\xi)H^2}\left(\frac{\partial^2 \Psi}{\partial \xi \partial \zeta} - \frac{H'}{H}\frac{\partial \Psi}{\partial \xi} - \frac{H'}{H}\frac{\partial^2 \Psi}{\partial \xi^2}(1+\xi)\right) \qquad (6.5)$$

$$\frac{\partial U}{\partial y} = \frac{8}{(1+\xi)H^3}\left(\frac{\partial^2 \Psi}{\partial \xi^2} - \frac{1}{(1+\xi)}\frac{\partial \Psi}{\partial \xi}\right) \qquad (6.6)$$

$$\frac{\partial V}{\partial z} = -\frac{2}{(1+\xi)H}\left\{\frac{\partial^2 \Psi}{\partial \zeta^2} - \left(\left(\frac{H''}{H} - \frac{2H'^2}{H^2}\right)\frac{\partial \Psi}{\partial \xi} + \frac{2H'}{H}\frac{\partial^2 \Psi}{\partial \zeta \partial \xi}\right)(1+\xi) + \frac{H'^2}{H^2}\frac{\partial^2 \Psi}{\partial \xi^2}(1+\xi)^2\right\} \qquad (6.7)$$

Note that $\partial V / \partial y$ is eliminated with the aid of the continuity equation, i.e., $\partial V / \partial y = -\partial U / \partial z - V / y$. As far as the initial conditions are concerned, the solution at the entrance region, i.e., for $z \leq 0$, are imposed for the stream function:

$$\Psi(y,0) \equiv \Psi_{en} = y^2(2 - y^2), \quad \frac{\partial \Psi}{\partial z}(y,0) \equiv \frac{\partial \Psi_{en}}{\partial z}(y,0) = 0 \qquad (6.8)$$

and, similarly, for the rescaled components of the conformation tensor (see § 4.1)

$$c_{zz}(y,0) = 128 De^2 \, y^2, \quad c_{yz}(y,0) = -8De\, y, \quad c_{yy}(y,0) = 1 \qquad (6.9)$$

We emphasize that the second condition given in Eq. (6.8b) requires special treatment in the new coordinate system (see Eq. (6.2) with $H(0) = 1$)

$$\frac{\partial \Psi}{\partial \zeta}(\xi,0) = H'(0^+)(1+\xi)\Psi'_{en}(\xi) \qquad (6.10)$$

The integration along $\zeta$ is performed by discretizing the computational domain in *N+1* equidistant grid points, $\zeta_n = n\,\Delta\zeta = n/N, \quad n = 0,1,2,3,\ldots,N$. We use the A-stable and first-order accurate finite-difference formula for $\partial f / \partial \zeta$

$$\frac{\partial f}{\partial \zeta}(\xi,\zeta_n) \equiv \left.\frac{\partial f}{\partial \zeta}\right|_{(n)} \approx \frac{f_{(n)} - f_{(n-1)}}{\Delta \zeta} \qquad (6.11a)$$

where $f = \Psi$, $c_{zz}$, $c_{yz}$ and $c_{yy}$, while for $\partial^2 \Psi / \partial \zeta^2$ we use the first-order accurate formula

$$\frac{\partial^2 \Psi}{\partial \zeta^2}(\xi,\zeta_n) \equiv \left.\frac{\partial^2 \Psi}{\partial \zeta^2}\right|_{(n)} \approx \frac{1}{\Delta \zeta}\left(\left.\frac{\partial \Psi}{\partial \zeta}\right|_{(n)} - \left.\frac{\partial \Psi}{\partial \zeta}\right|_{(n-1)}\right) \qquad (6.11b)$$

In the above formulas, $n \geq 1$ and a subscript in parenthesis denotes value at the corresponding grid $\zeta$-point, i.e., $f_{(n)} \equiv f(\xi,\zeta_n)$. This is achieved by developing a pseudospectral method. The field variables are given as series of Chebyshev orthogonal polynomials in terms of $\xi$ as follows

$$f_{(n)} \equiv f(\xi,\zeta_n) = \sum_{k=0}^{M} \hat{f}_{2k}(\zeta_n)T_k(\xi), \quad f = \Psi, c_{zz}, c_{yy}, c_{yz} \quad (6.12)$$

The Chebyshev orthogonal polynomials are defined as $T_k(\xi) = \cos(k\cos^{-1}(\xi))$, $k=0,1,2,...,M$ in the domain $\xi \in [-1,1]$. The spectral coefficients, namely the quantities with a hat in Eq. (6.12)) are calculated pseudospectrally based on the Gauss-Lobatto points (Hesthaven, Gottlieb & Gottlieb 2007) $\{\xi_j\}_{j=0}^{M} = \{-\cos(j\pi/M)\}_{j=0}^{M}$, at which the unknown physical values $f_{j,(n)} \approx f(\xi_j,\zeta_n)$ are assigned. The partial derivatives of $f$ with respect to $\xi$ at the Gauss-Lobatto points are evaluated effectively based on the differentiation matrices (Hesthaven, Gottlieb & Gottlieb 2007). Then, $f_{j,(n)}$ are calculated according to the discretized version of the governing equations and accompanied boundary conditions at the Gauss-Lobatto points. The resulting strongly non-linear system of algebraic equations is solved iteratively using a Newton scheme at each *n*-step yielding the solution for $f_{j,(n)}$; this scheme converges quadratically (as it should) with an absolute relative criterion $10^{-12}$ within two or three iterations.

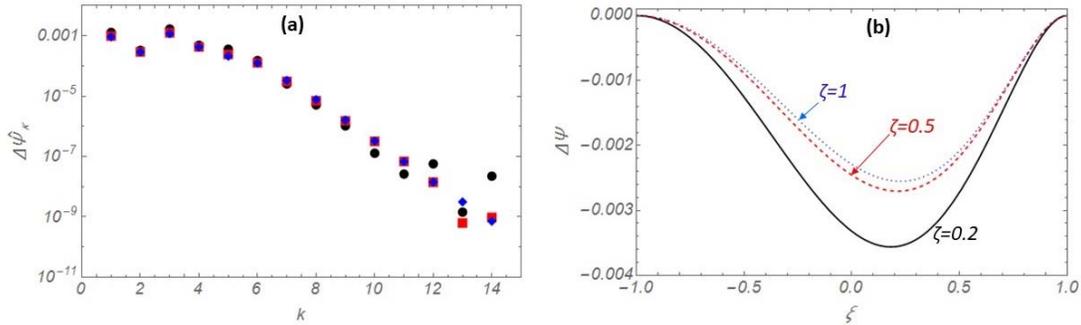

**Figure 7:** Simulations results for $\eta$ = 0.4 and $De_m$=0.45
(a) The magnitude of the spectral coefficients for $\Delta\Psi$; circles: $\zeta$=0.2, squares: $\zeta$=0.5; diamonds: $\zeta$=1
(b) $\Delta\Psi=\Psi-\Psi_N$ as function of the transformed radial coordinate $\xi$.

The accuracy of the solution calculated by our pseudospectral code is checked by monitoring the magnitude of the spectral coefficients for the stream function. In Figure 7a, $|\Delta\hat{\Psi}_k|$ is shown in logarithmic plot for *k*=1,2,…,*M*, where *M*=15, for $\zeta$=0.2, $\zeta$=0.5 (at the middle of the pipe) and $\zeta$=1 (at the outlet) for the case with $\eta = 0.4$ and $De_m = 0.45$. First, however, we report that $\Psi_{en}$ and $\Psi_N$ have the same form when are expressed in terms of the new coordinates

$$\Psi_{en}(\xi) = \Psi_N(\xi) = \frac{1}{16}(1+\xi)^2(7-2\xi-\xi^2) \tag{6.13}$$

as well as that there is no dependence on the $\zeta$ – coordinate. From Eq. (6.13) one can trivially confirm that $\Psi_j(-1) = \Psi'_j(-1) = \Psi'''_j(-1) = 0$, $\Psi_j(1) = 1, \Psi'_j(1) = 0$ where $j$="en" or "$N$", and also to calculate the corresponding spectral coefficients:

$$\left\{\hat{\Psi}_{en,k}\right\}_{k=0}^{4} = \left\{\hat{\Psi}_{N,k}\right\}_{k=0}^{4} = \left\{\frac{61}{128}, \frac{9}{16}, \frac{1}{32}, -\frac{1}{16}, -\frac{1}{128}\right\} \tag{6.14}$$

Figure 7a shows that the magnitude of the spectral coefficients of $\Delta\Psi \equiv \Psi(\xi,\zeta) - \Psi_j(\xi)$ drops fast, almost down to machine accuracy, revealing that the stream function is fully resolved. The very small magnitude of $\left|\Delta\hat{\Psi}_k\right|$, $k = 0,1,2,...,15$, reveals that the stream function, and consequently the velocity field, only slightly changes compared to the Newtonian velocity field (or compared to the solution for the straight pipe at the entrance region). The difference $\Delta\Psi$ is also shown as function of the transformed radial coordinate near the inlet ($\zeta$=0.2), at the middle of the pipe ($\zeta$=1/2) and at the outlet ($\zeta$=1) in Figure 7b.

In Table 2, we compare the results obtained with our pseudospectral method with the accelerated analytical solutions. The excellent agreement between the numerical and analytical results reveals the validity and accuracy for both the theoretical and numerical results. The same accuracy is also achieved for the individual contributions to the normalized pressure-drop (shown in Figure 4) and for the Trouton ratio (shown in Figure 6a).

| $De_m$ | 0.10 | 0.20 | 0.30 | 0.40 | 0.45 | 0.55 | 0.60 |
|---|---|---|---|---|---|---|---|
| $\Delta\Pi / \Delta\Pi_N$ | 0.9292 | 0.8730 | 0.8280 | 0.7916 | 0.7760 | 0.7491 | 0.7379 |
| $\Delta\Pi / \Delta\Pi_N\big|_{[1/1]}$ | 0.9289 (0.03%) | 0.8720 (0.12%) | 0.8255 (0.26%) | 0.7867 (0.62%) | 0.7696 (0.83%) | 0.7393 (1.32%) | 0.7257 (1.65%) |
| $\Delta\Pi / \Delta\Pi_N\big|_{[2/2]}$ | 0.9291 (0.01%) | 0.8730 (*) | 0.8281 (0.01%) | 0.7917 (0.01%) | 0.7760 (*) | 0.7488 (0.05%) | 0.7369 (0.13%) |
| $\Delta\Pi / \Delta\Pi_N\big|_{[3/3]}$ | 0.9291 (0.01%) | 0.8730 (*) | 0.8281 (0.01%) | 0.7916 (0.01%) | 0.7760 (*) | 0.7487 (0.05%) | 0.7368 (0.13%) |
| $\Delta\Pi / \Delta\Pi_N\big|_{[4/4]}$ | 0.9291 (0.01%) | 0.8730 (*) | 0.8281 (0.01%) | 0.7917 (0.01%) | 0.7760 (*) | 0.7487 (0.05%) | 0.7369 (0.13%) |

**Table 2:** Comparison of $\Delta\Pi / \Delta\Pi_N$ calculated numerically and analytically (rounded in four significant digits) using the Padé [M/M] diagonal approximants with M=1, 2, 3 and 4. The percentage relative absolute error is given in parenthesis; a star indicates error less than 0.01%

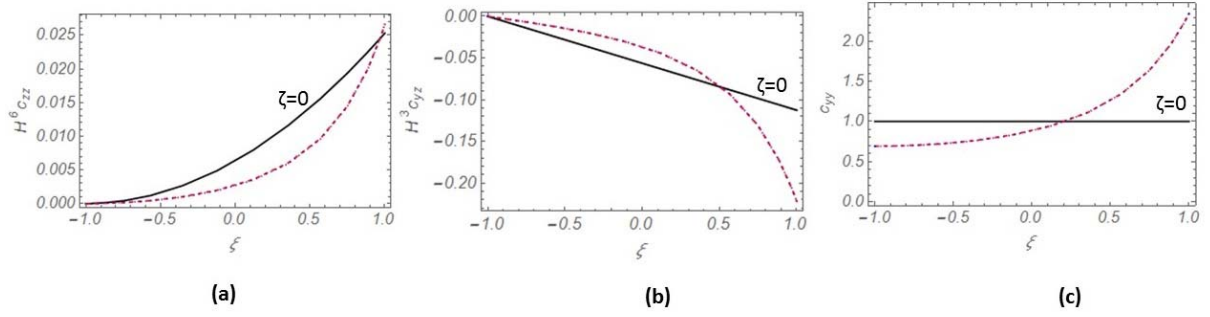

**Figure 8:** Simulations results for Λ=3 and $De_m$=0.45 as function of the transformed radial coordinate, $\xi$.
(a) $H^6 c_{zz}$ (b) $H^3 c_{yz}$ (c) $c_{yy}$
All curves for $\zeta > 0$ (shown with dashed lines) collapse

Last, in Figure 8, we present results for the rescaled components of the conformation tensor, $H^6 c_{zz}$, $H^3 c_{yz}$ and $c_{yy}$ as functions of the transformed radial coordinate, $\xi$. It is seen that these quantities remain practically the same for any ζ>0; the curves for ζ=0 correspond to the initial condition at the inlet. The results reveal that $c_{yz} \approx c_{yz,N}$, $c_{yy} \approx c_{yy,N}$ and $c_{zz} \gg c_{zz,N} = 0$ which clearly indicates that $c_{zz}$ (and the corresponding extra-stress $\tau_{zz}$) is the leading contribution to the pressure drop due to the viscoelasticity of the fluid.

## 7. Discussion

We showed above that the high-order asymptotic formula for the Trouton ratio, Eq. (5.25), at the axis of symmetry requires special attention. The main reason is that for the hyperbolic geometry the normal component of the conformation tensor, $c_{zz}$, (or $\tau_{zz}$,) shows a peculiar behavior at $De_m$=1/2; consequently, this peculiarity affects the Trouton ratio at $De_m$=1/2 too. Thus, it did not come as a surprise that the perturbation solution and the corresponding transformed formulas diverge near that point, which clearly poses an upper bound for their validity. These results are closely related to the known deficiency of the Oldroyd-B model which predicts an infinite Trouton ratio in homogeneous uniaxial extensional flow at a finite extension rate which corresponds to $De_m = 1/2$ (Bird, Armstrong & Hassanger 1987; Tanner 2000; Housiadas 2017).

The explanation for this peculiar behavior of the Trouton ratio at $De_m = 1/2$ is given with the aid of the exact analytical solution for $C_{zz}(0,z)$ (see Eq. (4.16)) and its limit as $De_m \to 1/2$ given by Eq. (4.18). These solutions directly affect the Trouton ratio at the exit of the pipe and its limit as $De_m \to 1/2$ given by Eqs. (4.20) and (4.21), respectively. It is important to mention here that the exact solutions for $C_{zz}$ and $Tr$ at the axis of symmetry are well-behaved for any $De_m > 0$. The formula around which the exact solution for $Tr$ was developed, Eq. (3.18b), reveals the presence of a power law term with exponent proportional to the inverse of the original Deborah number (see Eq. (3.16)). This term is not amenable to a representation in terms of a regular perturbation series with respect to $De$, causing the series to diverge at infinity at the same modified Deborah number as the exact solution for pure homogeneous uniaxial extensional flow (i.e., at $De_m = 1/2$).

One can therefore naturally ask the question: suppose we do not know the well-behaved for any $De_m > 0$ exact solution. Is it possible, using a regular perturbation scheme, to provide reasonable and/or accurate predictions for the Trouton ratio at $De_m$ values larger than 1/2? Although a strict and formal answer to this important issue cannot be given, previous work by Housiadas (2017) on pure homogeneous elongational flows showed that using only a series developed asymptotically as $De_m \to 0^+$, this is not possible. Even worse, the use of more realistic (and highly non-linear) models like the Giesekus, FENE-P (Bird, Armstrong & Hassanger 1987) and Phan-Thien & Tanner (Tanner 2000) models does not circumvent or resolve this issue.

However, recent work by Housiadas (2023) revealed that a possible answer to this question is to consider both limits (low and high Deborah numbers), in deriving asymptotic solutions. Then, use of convergence acceleration methods (like, for instance, two-point Padé approximants) followed by the development of uniformly valid approximations, can solve this problem. The performance of this method, which depends on the available number of analytical terms at both limits, can be amazing. Another possible answer to this question may be to consider advanced perturbation schemes, such as the exponential asymptotics presented by Kataoka & Akylas 2022, that may capture more efficiently the anomaly/peculiar behavior of the constitutive model at $De_m$=1/2. This issue requires further in-depth investigation because for a long time is believed that the regular perturbation schemes in terms of the Deborah or the Weissenberg numbers are the only ones that should be used for developing asymptotic solutions for weakly viscoelastic flows (Bird, Armstrong & Hassanger 1987).

## 8. Conclusions

We studied theoretically the steady viscoelastic flow of an Oldroyd-B fluid under creeping conditions in a long axisymmetric tube with variable cross-section, focusing on the hyperbolic contracting pipe. First, we developed the general theoretical framework for the evaluation of the average pressure drop, required to maintain a constant flow rate through the pipe. Second, we proved that the velocity field along the axis of symmetry of the pipe is a pure uniaxial extensional field, and we found the exact analytical solution for the components of the constitutive model at the axis of symmetry in terms of the fluid velocity only. The solution for the constitutive model along with the observation that the flow velocity changes little with viscoelasticity along the axis of symmetry (at least in the limit of a small aspect ratio) allowed for a robust and accurate evaluation of the first normal stress difference up to high Deborah numbers, through which the dimensionless elongational viscosity of the fluid (or Trouton ratio) was extracted. For a hyperbolic contracting pipe and approximating the velocity with that for a Newtonian fluid under the same conditions and geometry, we showed that the Trouton ratio increases with increasing Deborah number, polymer viscosity ratio, and contraction ratio.

Further analytical progress was achieved by invoking the classic lubrication approximation and solving the final equations for an arbitrary shape function, using a high-order asymptotic scheme in terms of the original Deborah number, *De*. In this case, the entrance and exit regions of the tube do not play any role in the analysis, and consequently they do not affect the results. The lubrication equations were solved analytically up to eight-order in *De*, resulting in analytical formulas for all the field variables and in particular for the average pressure drop, $\Delta\Pi$, and the Trouton ratio, $Tr$. For the hyperbolic pipe, we revealed that the high-order perturbation solutions, when reduced by their corresponding Newtonian values, namely the quantities $\Delta\Pi/\Delta\Pi_N$ and $Tr$ given by Eqs. (5.11) and (5.25), respectively, can be recast in terms of a modified Deborah number, $De_m$, and the polymer viscosity ratio, $\eta$, only. The convergence of the perturbation series was checked and enhanced by deriving transformed solutions using Padé approximants. The latter showed convergence for $\Delta\Pi/\Delta\Pi_N$ for any $De_m > 0$, revealing a monotonic decrease with increasing $De_m$ and/or $\eta$. On the contrary, the transformed solutions for $Tr$ converged only in the window $0 < De_m < 1/2$, showing an increase of the Trouton ratio with increasing $De_m$ and/or $\eta$, and diverging near $De_m \approx 1/2$.

This deficiency/divergence was traced to the very peculiar dependence of the exact analytical solution for the Trouton ratio on $De_m$ and $z$ which changes continuously from an algebraic type dependence for $0 < De_m < 1/2$, to a logarithmic type dependence at $De_m = 1/2$, and back to the algebraic type for $De_m > 1/2$. Therefore, the transformed perturbation solution(s) for the Trouton ratio can be used only for $0 < De_m < 1/2$ while the new developed exact formula (Eq. (3.18a) or Eq. (3.18b)) allow us to get well-behaved and accurate predictions even for $De_m \geq 1/2$ given of course that a good approximation for the fluid velocity is available. We were also able to show that even the Newtonian velocity profile can give a sufficient result, which fully justifies this approach followed by many researchers in the literature. Thus, the corresponding formula for the Trouton ratio given by Eq. (4.18), valid for a long hyperbolic axisymmetric pipe, represents the best theoretical prediction for Boger-type fluids (i.e. viscoelastic fluids with negligible shear thinning behavior). Moreover, comparison of Eq. (4.18) with the high-order perturbation solution for $Tr$, post-processed with the Padé diagonal transformation, revealed that the two formulas agree very well in the range $0 < De_m < 0.3$ approximately. Note that these observations and predictions are in full accordance with the well-known singularity of the Oldroyd-B model in homogeneous uniaxial extensional flows at a finite rate of extension which corresponds to $De_m = 1/2$.

Finally, we mention that our theoretical predictions, the analytical formulas, as well as their transformed solutions were checked for their accuracy and validity through comparison with numerical results obtained using pseudospectral methods to solve the lubrication equations. The comparison revealed excellent agreement between the solutions clearly demonstrating the accuracy, robustness and suitability of the theoretical methods and techniques used here.

**Declaration of Interests**

The authors report no conflict of interest.

**Appendix**

Functions $\hat{\Psi}_2^{(2k)} = \hat{\Psi}_2^{(2k)}(z)$, $k=1,2,3,4$ are:

$$\hat{\Psi}_2^{(2)} = \frac{64(HH'' - 4H'^2)}{3H^6}, \quad \hat{\Psi}_2^{(4)} = \frac{176H'^2 - 48HH''}{H^6},$$

$$\hat{\Psi}_2^{(6)} = \frac{32(HH'' - 3H'^2)}{H^6}, \quad \hat{\Psi}_2^{(8)} = \frac{16(H'^2 - HH'')}{3H^6}$$

(A1)

Functions $\hat{\tau}_{yz,2}^{(2k+1)} = \hat{\tau}_{yz,2}^{(2k+1)}(z)$, $k=0,1,2$ are:

$$\hat{\tau}_{yz,2}^{(1)} = -15H'^2 + 3HH'' + \frac{\eta}{2}\left(11H'^2 - 3HH''\right)$$

$$\hat{\tau}_{yz,2}^{(3)} = 42H'^2 - 6HH'' + 3\eta\left(HH'' - 3H'^2\right)$$

$$\hat{\tau}_{yz,2}^{(5)} = -27H'^2 + 3HH'' + \eta\left(H'^2 - HH''\right)$$

(A2)

Functions $\hat{\tau}_{yy,2}^{(2k)} = \hat{\tau}_{yy,2}^{(2k)}(z)$, $k=0,1,2,3$ are:

$$\hat{\tau}_{yy,2}^{(0)} = 40H'^3 - 17HH'H'' + H^2H''' + \eta\left(-\frac{32}{3}H'^3 + 5HH'H'' - \frac{1}{3}H^2H'''\right)$$

$$\hat{\tau}_{yy,2}^{(2)} = -180H'^3 + 63HH'H'' - 3H^2H''' + \eta\left(\frac{165H'^3}{2} - \frac{147}{4}HH'H'' + \frac{9}{4}H^2H'''\right)$$

$$\hat{\tau}_{yy,2}^{(4)} = 264H'^3 - 75HH'H'' + 3H^2H''' + \eta\left(-90H'^3 + \frac{85}{2}HH'H'' - \frac{5}{2}H^2H'''\right)$$

$$\hat{\tau}_{yy,2}^{(6)} = -124H'^3 + 29HH'H'' - H^2H''' + \eta\left(\frac{49H'^3}{6} - \frac{35}{4}HH'H'' + \frac{7}{12}H^2H'''\right)$$

(A3)

Functions $\hat{\tau}_{\theta\theta,2}^{(2k)} = \hat{\tau}_{\theta\theta,2}^{(2k)}(z)$, $k=0,1,2,3$ are:

$$\hat{\tau}_{\theta\theta,2}^{(0)} = 40H'^3 - 17HH'H'' + H^2H''' + \eta\left(-\frac{32}{3}H'^3 + 5HH'H'' - \frac{1}{3}H^2H'''\right)$$

$$\hat{\tau}_{\theta\theta,2}^{(2)} = -120H'^3 + 51HH'H'' - 3H^2H''' + \eta\left(\frac{55H'^3}{2} - \frac{49}{4}HH'H'' + \frac{3}{4}H^2H'''\right)$$

$$\hat{\tau}_{\theta\theta,2}^{(4)} = 120H'^3 - 51HH'H'' + 3H^2H''' + \eta\left(-18H'^3 + \frac{17}{2}HH'H'' - \frac{1}{2}H^2H'''\right)$$

$$\hat{\tau}_{\theta\theta,2}^{(6)} = -40H'^3 + 17HH'H'' - H^2H''' + \eta\left(\frac{7H'^3}{6} - \frac{5}{4}HH'H'' + \frac{1}{12}H^2H'''\right)$$

(A4)